\documentclass{ptephy_v1}

\usepackage{amsmath}
\usepackage{amsfonts}
\usepackage{amssymb}
\usepackage{ulem}
\usepackage{eucal}

\usepackage{color}
\usepackage{graphicx}

\definecolor{red  }{rgb}{1,0,0}
\definecolor{blue }{rgb}{0,0,1}
\definecolor{green}{rgb}{0,1,0}
\definecolor{CiteColor}{rgb}{0,0,0.35}
\definecolor{URLColor}{rgb}{0,0,0.35}
\definecolor{darkgreen}{rgb}{0.2,0.7,0.2}

\newcommand{\rmd}{{\mathrm {d}}}
\newcommand{\Hint}{H^{(1)}}
\newcommand{\Hrad}{{H}_{\mathrm {rad}}}
\newcommand{\Hsym}{{H}_{\mathrm {sym}}}

\newcommand{\conj}[1]{\overline{#1}}

\newcommand{\wlm}{{\omega \ell m}}
\newcommand{\wlmkn}{{\omega \ell m k n}}

\allowdisplaybreaks
\begin{document}

\title{``Flux-balance formulae'' for extreme mass-ratio inspirals}

\author{\name{Soichiro Isoyama}{1,2},
\name{Ryuichi Fujita}{3}, 
\name{Hiroyuki Nakano}{4},
\name{Norichika Sago}{5}, 
and 
\name{Takahiro Tanaka}{3,6}}

\address{
${}^1$\affil{1}{Department of Physics, University of Guelph, Guelph, 
Ontario, N1G 2W1, Canada}
\\
${}^2$\affil{2}{International Institute of Physics, 
Universidade Federal do Rio Grande do Norte, Campus Universitario, Lagoa Nova, 
Natal-RN 59078-970, Brazil}
\\
${}^3$\affil{3}{Center for Gravitational Physics, Yukawa Institute 
for Theoretical Physics, Kyoto University,Kyoto 606-8502, Japan}
\\
${}^4$\affil{4}{Faculty of Law, Ryukoku University, Kyoto 612-8577, Japan}
\\
${}^5$\affil{5}{Faculty of Arts and Science, Kyushu University, 
Fukuoka 819-0395, Japan}
\\
${}^6$\affil{6}{Department of Physics, Kyoto University, 
Kyoto 606-8502, Japan}}


\begin{abstract}

The ``flux-balance formulae'' that determine the averaged evolution of energy, azimuthal angular momentum, and Carter constant in terms of the averaged asymptotic gravitational-wave fluxes for inspirals of small bodies into Kerr black holes were first derived about 15 years ago. However, this derivation is restricted to the case that the background Kerr geodesics are non-resonant (i.e., the radial and angular motions are always incommensurate), and excludes the resonant case that can be important for the radiative dynamics of extreme mass-ratio inspirals. We give here a new derivation of the flux formulae based on Hamiltonian dynamics of a self-forced particle motion, which is a valuable tool for analyzing self-force effects on generic (eccentric, inclined) bound orbits in the Kerr spacetime. This Hamiltonian derivation using action-angle variables is much simpler than the previous one, applies to resonant inspirals without any complication, and can be straightforwardly implemented by using analytical/numerical Teukolsky-based flux codes. 

\end{abstract}




\subjectindex{E01, E02, E31, E38}

\maketitle

\section{Introduction and summary}
\label{sec:intro}

Consider an extreme mass-ratio inspiral of a smaller non-spinning body 
of mass $\mu$ into a larger Kerr black hole of mass $M \gg \mu$, 
which is a main target of the space-based 
gravitational wave observatory of
LISA~\cite{Audley:2017drz,Armano:2018kix,Babak:2017tow}.
In the case of $\eta \equiv \mu / M \ll 1$ 
this two-body problem lends itself to a systematic perturbative method, 
now known as ``self-force theory.'' 
In the test mass limit $\eta \to 0$, 
the body's orbits reduce to the timelike geodesic orbits 
in the background Kerr spacetime. 
At subsequent $O(\eta)$, interaction of the body 
with its own gravitational field causes a ``gravitational self-force'' (GSF), 
and the self-force drives the slow inspirals accompanied by
the gravitational radiation. Reference~\cite{Barack:2018yvs} is 
the most recent colloquium-style review of the self-force theory including
the latest achievements, 
and the expert-level reviews of Refs.~\cite{Mino:2005yw,Tanaka:2005ue,
Barack:2009ux,Poisson:2011nh,Harte:2014wya,Pound:2015tma}
follow up the technical details. 

When modeling the orbital dynamics of extreme mass-ratio inspirals, 
it is usually convenient to track the evolution of conserved quantities
along the background geodesics 
(constants of motion, orbital frequencies, etc.) if the background spacetime 
has sufficiently many symmetries. 
This is the so-called ``adiabatic description'' of extreme mass-ratio 
inspirals: see Sec.~5 of Ref.~\cite{Barack:2018yvs} for reviews. 
It is known that timelike geodesics of the Kerr geometry are integrable 
and admit three non-trivial constants of motion associated with 
Killing fields of the Kerr geometry~\cite{Carter:1968rr}: 
the (specific) energy ${\hat E}$ and azimuthal angular momentum ${\hat L}$ 
defined by the Killing vectors, and the \textit{Carter constant} ${\hat Q}$ 
defined by the second-rank {Killing tensor} 
(see Appendix~\ref{sec:Killing}). 
Furthermore, generically, it is known that bound geodesic motions
in the Kerr spacetime are tri-periodic with the trio of frequencies 
$\{\Omega^{r},\,\Omega^{\theta},\,\Omega^{\phi}\}$ 
in terms of the Boyer-Lindquist time coordinate 
$t$~\cite{Schmidt:2002qk}~\footnote{
The spherical (equatorial) orbits are biperiodic motions  
because $\Omega^{r}(\Omega^{\theta})$ loses its meaning from the first place.
}---hereafter, $(t,r,\theta,\phi)$ are the Boyer-Lindquist coordinates 
and we will work in the geometrized units $G = c = 1$ 
with the $(-,\,+,\,+,\,+)$ metric signature. 

At $O(\eta)$, both triplets of $\{{\hat E},\,{\hat L},\,{\hat Q}\}$ 
and $\{\Omega^{r},\,\Omega^{\theta},\,\Omega^{\phi}\}$ 
evolve in time under the self-force effects. 
Nevertheless, the leading-order evolution of the orbital phase, 
which is the most important information that we need 
for the gravitational-wave applications 
(e.g., Refs.~\cite{Barack:2003fp,Lindblom:2008cm}), 
only requires the time-averaged rates of change 
of $\{{\hat E},\,{\hat L},\,{\hat Q}\}$ 
as input~\cite{Pound:2007ti,Hinderer:2008dm}. 
Such averaged rates of change are easily expressed 
in terms of the $O(\eta)$ self-force
(e.g., Ref.~\cite{Ori:1997be}; see also Appendix~\ref{sec:dQdt-GSF}), 
but practical challenges enter here 
because the computational methods of self-force in the Kerr spacetime 
are expensive and technically challenging~\cite{Wardell:2015kea,vandeMeent:2016pee,vandeMeent:2017bcc}. 
Therefore, it is useful to have an alternative expression 
that allows us to compute the rates of change of 
$\{{\hat E},\,{\hat L},\,{\hat Q}\}$ 
without resorting to the actual (numerical) self-force calculation. 

A natural strategy would be to apply the ``flux-balance formulae'' 
that determine the time-averaged rates of change 
of $\{{\hat E},\,{\hat L},\,{\hat Q}\}$ under the self-force 
(``radiation-reaction force'') effects from the associated 
total gravitational radiation out to infinity 
and down to the horizon of the Kerr black hole: 
see Sec.~6 of Ref.~\cite{Barack:2018yvs} for reviews.  
Given that computation of asymptotic gravitational-wave fluxes 
has been routinely performed since the 1970s 
(see, e.g., Refs.~\cite{Nakamura:1987zz,Sasaki:2003xr} 
for reviews, and visit the repositories of Refs.~\cite{BHPC, BHPT}), 
the flux formulae enable highly accurate and efficient computation 
of the averaged rates of change of $\{{\hat E},\,{\hat L},\,{\hat Q}\}$. 
Because the source orbit of the self-force at $O(\eta)$ 
can be approximated by the bound Kerr geodesic in the leading-order 
adiabatic description, there are two cases to consider: 
(i) the source orbit is a \textit{non-resonant} one, i.e.,
the radial and polar motions of the Kerr geodesic are incommensurate;  
(ii) the source orbit is a \textit{resonant} one, i.e., 
the ratio of orbital frequencies $\Omega^{r}/\Omega^{\theta}$ 
becomes a rational number and hence degenerate~\cite{Mino:2005an,
Tanaka:2005ue,Flanagan:2010cd,Brink:2013nna,Brink:2015roa}. 
It is now known that every extreme mass-ratio inspiral 
that will be observable by LISA is expected to pass through at least one resonance~\cite{Ruangsri:2013hra}, and neglecting this effect would lead 
to a loss of detectable gravitational-wave signals~\cite{Berry:2016bit}; 
see Sec.~5.3 of Ref.~\cite{Barack:2018yvs} 
as well as Refs.~\cite{Gair:2011mr,Mihaylov:2017qwn} 
for the further implications of this $r\mathchar`-\theta$ resonance 
for gravitational-wave astronomy.~\footnote{
There are different types of orbital resonance: 
$r\mathchar`-\phi$ resonance~\cite{vandeMeent:2014raa}
and $\theta\mathchar`-\phi$ resonance~\cite{Hirata:2010xn}, 
where the ratios of $\Omega^{r}/\Omega^{\phi}$ 
and $\Omega^{\theta}/\Omega^{\phi}$ become rational numbers, respectively. 
However, these resonances can be handled 
by a subset of ``non-resonant'' cases when deriving the flux formulae, 
and we are not concerned here with these resonances.} 

In the non-resonant case (i), the flux formulae were established  
by Sago et al.~\cite{Sago:2005gd,Sago:2005fn,Ganz:2007rf} 
and Drasco et al.~\cite{Hughes:2005qb,Drasco:2005is,Drasco:2005kz} 
almost 15 years ago. The critical step in their derivation relies 
on a breakthrough result by Mino~\cite{Mino:2003yg}, 
which is based on the early work by Gal'tsov~\cite{Galtsov:1982hwm}.    
Mino showed that the time-averaged rates of change 
of $\{{\hat E},\,{\hat L},\,{\hat Q}\}$ for non-resonant orbits 
are driven only by that of the dissipative piece of the self-force 
constructed from the (half-retarded-minus-half-advanced) 
radiative gravitational perturbation. 
The radiative perturbation is a global vacuum solution 
of the linearized Einstein equation, 
and this allows one to rewrite the time-averaged radiative self-force 
in terms of the asymptotic amplitudes 
of gravitational radiation at infinity and the horizon. 
The flux formulae then follow from the time-averaged 
self-forced equations of motion. 
The time-averaged rates of change of ${\hat E}$ and ${\hat L}$ 
calculated by the flux formulae are matched precisely 
by the total time-averaged  fluxes of 
energy and angular momentum out to infinity and down to the horizon 
of a Kerr black hole, as expected~\cite{Sasaki:2003xr,Drasco:2005kz}. 
Although there are no known gravitational-wave fluxes 
of Carter constant ${\hat Q}$ 
(see, e.g., Ref.~\cite{Grant:2015xqa}), somewhat surprisingly, 
this self-force derivation can provide the desired flux formulae 
for the time-averaged rates of change of ${\hat Q}$. 

However, the generalization of the flux formulae to the resonant case (ii) 
does not appear straightforward~\footnote{
The time-averaged asymptotic fluxes of energy and angular momentum 
from the resonant orbit have recently been computed by 
Grossman, Levin, and Perez-Giz~\cite{Grossman:2011im}, 
and Flanagan, Hughes, and Ruangsri~\cite{Flanagan:2012kg}.
}.
The technical difficulties and subtleties trace back 
to the dependence on initial phases 
of the orbits in the resonant case~\cite{Brink:2015roa},
and in fact Refs.~\cite{Mino:2003yg,Sago:2005gd,Sago:2005fn}
take full advantage of non-resonant assumptions
to prove the absence of this initial phase dependence.
The non-resonant assumptions in the analysis by Mino~\cite{Mino:2003yg} 
and Sago et al.~\cite{Sago:2005gd,Sago:2005fn} have been partially removed 
by Hinderer and Flanagan~\cite{Hinderer:2008dm,Flanagan:2010cd}
as well as the present authors~\cite{Isoyama:2013yor,Fujita:2016igj} 
and Flanagan, Hughes, and Ruangsri~\cite{Flanagan:2012kg}, respectively.  
The purpose of this paper is to eliminate 
the remaining non-resonant restrictions, 
and to establish the complete set of flux formulae for radiative inspirals 
of small bodies into Kerr black holes at $O(\eta)$, 
including the case of resonant orbits.

\subsection{Outline and summary of this paper}

Our approach to the flux formulae is based 
on a Hamiltonian approach in the self-force theory, 
originally developed for the conservative self-force dynamics 
in the Kerr spacetime~\cite{Isoyama:2014mja, Vines:2015efa, Fujita:2016igj}. 
The Hamiltonian method allows us to conveniently formulate 
the (first-order) self-forced equations of motion  
as the Hamilton's canonical equations in terms of action-angle variables 
on the phase space~\cite{Schmidt:2002qk,Hinderer:2008dm}, 
which are fully separable in the test-mass limit 
because the action-angle variables manifestly respect 
the integrability and tri-periodicity of the bound Kerr geodesic orbits.
Furthermore, all the self-force effects are encoded in a single 
(two-body) interaction piece of the Hamiltonian with efficient applications 
of Hamiltonian frameworks in mathematical physics. 
As a consequence, the analysis has a great advantage over the previous 
non-systematic spacetime approach of Refs.~\cite{Mino:2003yg,Sago:2005gd,Sago:2005fn,Drasco:2005is,Isoyama:2013yor} to unify the flux formulae in both non-resonant and resonant cases.

We begin our derivation in Sec.~\ref{sec:Hamilton} 
with a review of our Hamiltonian formalism of self-force dynamics 
in the Kerr geometry~\cite{Isoyama:2014mja, Fujita:2016igj}. 
Based on the generalized equivalence principle 
(i.e., the self-forced motion in the background Kerr spacetime
is equivalent to geodesic motion in a certainly perturbed Kerr
spacetime)~\cite{Detweiler:2002mi,Pound:2012nt}, 
the Hamiltonian method describes the motion
of a nonspinning point particle by the geodesic Hamiltonian, 
$H \equiv H^{(0)} + \Hint$
on the $8$D phase space spanned by the canonical variables 
$(x^{\mu},\,u_{\mu})$ with the effective metric
$g_{\mu \nu} \equiv g_{\mu \nu}^{(0)} + h_{\mu \nu}^{(\text{R})}$. 
Here, $H^{(0)}$ is the background Hamiltonian 
for timelike geodesics in the background Kerr metric $g_{\mu \nu}^{(0)}$, 
and $H^{(1)} (\propto \eta)$ is an interaction Hamiltonian that 
accounts for the (first-order) self-force generated by the ``regular'' part 
of the metric perturbation $h_{\mu \nu}^{(\text{R})} (\propto \eta)$ 
associated with the orbit. 
We work out the Hamiltonian $H$ in terms of 
the action-angle variables $(w^{\alpha},\, J_{\alpha})$ 
defined in Eqs~\eqref{def-J0} and~\eqref{def-w}, 
and the Hamilton equation for the actions $J_{\alpha}$ 
is given by   
\begin{equation}\label{Heq-wJ0}
{\dot J}_{\alpha} 
 = 
 - \left(\frac{\partial H^{(1)}}{\partial w^{\alpha}}\right)_{\!J}  \,, 
\end{equation}
where the overdot denotes the derivative with respect to 
the proper time $\tau$ along the orbit (measured in $g_{\mu \nu}$).  
Notice that ${\dot J}_{\alpha}$ involves only the interaction Hamiltonian 
$H^{(1)}$ because the background Hamiltonian is $H^{(0)} = H^{(0)}(J)$.  

In Sec.~\ref{sec:rad-sym}, we simplify the Hamilton equation 
of Eq.~\eqref{Heq-wJ0} by the long-time average. 
We split the interaction Hamiltonian $H^{(1)}$ 
into time-antisymmetric (``radiative'') pieces $\Hrad$ 
and time-symmetric piece $\Hsym$,\footnote{
In standard GSF language, $\Hrad(\Hsym)$ corresponds 
to the dissipative (conservative) piece of the self-force.}
i.e.,~$H^{(1)} = \Hrad + \Hsym$, and show that 
the general (quasi-gauge-invariant) expression 
for the averaged Hamilton's equation of ${\dot J}_{\alpha}$
is~\footnote{The symmetric contribution to Eq.~\eqref{Jdot0} 
in the resonant case was first pointed out 
using a scalar-field toy model---see Eq.~(59) of Ref.~\cite{Isoyama:2013yor}.}
\begin{equation}\label{Jdot0}
\langle \dot J_{\alpha} \rangle_\tau
= 
-\left\langle \left(
\frac{\partial \Hrad}{\partial w^{\alpha}} \right)_{\!J}
\right\rangle_\tau 
- \frac{1}{2} (\delta_{\alpha}^{r} +  \delta_{\alpha}^{\theta} )
\frac{\delta}{\delta w^{\alpha}_0}
\left\langle  \Hsym \right\rangle_\tau \,,
\end{equation}
where $\langle \cdot \rangle_{\tau}$ is the average over 
the proper time $\tau$, 
and ${\delta}/{\delta w^{\alpha}_0}$ is the total variation 
with respect to the initial values of angle variables $w^{\alpha}_0$. 
This equation applies to both the non-resonant and resonant orbits, 
but we will show that the non-resonant orbit yields
${\delta \langle \Hsym \rangle_\tau} / {\delta w^{\alpha}_0} = 0$ 
because $\langle  \Hsym \rangle_\tau$ is guaranteed 
to be independent of $w^{\alpha}_0$ in this case. 

In Sec.~\ref{sec:flux}, we establish a connection 
between the $\tau$-averaged Hamilton equation of Eq.~\eqref{Jdot0} 
and the amplitudes of the gravitational radiation 
at future null infinity and the horizon. 
The basic idea is to derive the mode expression of $\Hrad$ 
based on the Teukolsky formalism; 
the (half-retarded-minus-half-advanced) 
radiative metric perturbation contained in $\Hrad$ is 
a {global} vacuum perturbation, whose asymptotic forms can be expressed 
in terms of mode solutions to the homogeneous Teukolsky equation. 
We then show our main results of flux formulae for ${J}_{\alpha}$. 

In the case of the non-resonant orbit, the flux formulae are expressed, 
see Eq.~\eqref{Flux-nonres}, as 
\begin{equation}\label{dJdt-nonres0}
\left \langle 
\frac{\rmd J_{\alpha}}{\rmd t} 
\right \rangle_t 
=
- \mu \sum_{\ell m k n} 
\frac{{\varepsilon}_{\alpha}}{4 \pi \omega^3_{m k n}}
\left(
| {\tilde Z}_{\ell m k n}^{\mathrm {out}} |^2
+
\frac{\omega_{m k n}}{p_{m k n}} 
| {\tilde Z}_{\ell m k n}^{\mathrm {down}} |^2 
\right)\,, 
\end{equation}
where $\langle \cdot \rangle_t$ is the average 
over an asymptotic time coordinate $t$, 
$\ell$ is the (spheroidal) angular number of gravitational wave modes, 
$\omega_{m k n}$ and $p_{m k n}$ are the functions 
of the linear combination of the orbital frequencies $\Omega^{\alpha}$ 
in terms of mode integers $\{m,\,k,\,n \}$, 
${\varepsilon}_{\alpha} \equiv (-\omega_{m k n},\,n,\,k,\,m)$, 
and 
${\tilde Z}_{\ell m k n}^{\mathrm {out/dowm}}$ are asymptotic amplitudes 
of gravitational waves  modes ``out (down)'' to infinity (the horizon). 

In the case of the resonant orbits 
---see Eqs.~\eqref{Flux-res} and~\eqref{dJbdt-res}--- 
we have instead 
\begin{subequations}\label{dJdt-res0}
\begin{align}
\left \langle 
\frac{\rmd J_{A}}{\rmd t} 
\right \rangle_t 
&=
- \mu \sum_{\ell m N} 
\frac{{\varepsilon}_{A}}{4 \pi \omega^3_{m N}}
\left(
| {\tilde {\cal Z}}_{\ell m N}^{\mathrm {out}} |^2
+
\frac{\omega_{m N}}{p_{m N}} 
| {\tilde {\cal Z}}_{\ell m N}^{\mathrm {down}} |^2 
\right)
\quad \text{for $A \equiv (t,\, \phi)$}\,, 
\\
\left \langle 
\frac{\rmd J_{\parallel}}{\rmd t} 
\right \rangle_t 
&=
- \mu \sum_{\ell m N} 
\frac{N}{4 \pi \omega^3_{m N}}
\left(
| {\tilde {\cal Z}}_{\ell m N}^{\mathrm {out}} |^2
+
\frac{\omega_{m N}}{p_{m N}} 
| {\tilde {\cal Z}}_{\ell m N}^{\mathrm {down}} |^2 
\right)\,,
\end{align}
\end{subequations}
in which 
${\varepsilon}_{A} \equiv (-\omega_{m N},\,m)$ with a mode integer $N$, 
and 
$\langle {\rmd {J}_{\parallel}}/{\rmd t} \rangle_t 
\equiv 
\beta^{r} \langle {\rmd J_{r}}/{\rmd t} \rangle_t 
+
\beta^{\theta} \langle {\rmd J_{\theta}}/{\rmd t} \rangle_t$ 
with integers $\beta^{r}$ and $\beta^{\theta}$ that characterize resonance 
(i.e., $\Omega^{r}/\Omega^{\theta} = \beta^{r}/\beta^{\theta}$). 
Here, the on-resonance ``out/down'' amplitudes of gravitational wave modes 
${\tilde {\cal Z}}_{\ell m N}^{\mathrm {out/down}}$
explicitly depend on the initial value $w^{\alpha}_0$ of the orbit 
(see Eqs.~\eqref{cal-XYZ}). 
Importantly, to evaluate $\langle {\rmd J_{r}}/{\rmd t} \rangle_t$ 
or $\langle {\rmd J_{\theta}}/{\rmd t} \rangle_t$, we need 
the additional symmetric contribution of 
${\delta \langle \Hsym \rangle_\tau} / {\delta w^{\alpha}_0}$
as well as $\langle {\rmd {J}_{\parallel}}/{\rmd t} \rangle_t$: 
recall Eq.~\eqref{Jdot0}.

In Sec.~\ref{sec:dPdt} we derive the flux formulae 
for the $t$-averaged rates of change of specific energy ${\hat E}$, 
azimuthal angular momentum ${\hat L}$ and {Carter constant} ${\hat Q}$, 
built on the flux formulae of Eqs.~\eqref{dJdt-nonres0} 
and~\eqref{dJdt-res0}.   
In the Hamiltonian formalism, 
the triplets of $\{{\hat E},\,{\hat L},\,{\hat Q}\}$ 
are canonical variables on the phase space, 
which are invertible functions of $J_\alpha$. 
After a simple canonical transformation, we obtain 
$
\langle {\rmd {\hat E}}/{\rmd t} \rangle_t 
= 
- \langle {\rmd J_{t}}/{\rmd t} \rangle_t,\, 
\langle {\rmd {\hat L}}/{\rmd t} \rangle_t 
= 
\langle {\rmd J_{\phi}}/{\rmd t} \rangle_t
$, and, (see Eqs.~\eqref{dotQC-ave}) 
\begin{equation}\label{dotQ-ave0}
\frac{1}{2}
\left \langle \frac{\rmd {\hat Q}}{\rmd t} \right \rangle_t 
=- 
\left \langle \frac{r^2 + a^2}{\Delta} {\hat P} \right \rangle_\lambda
\left \langle \frac{\rmd J_t}{\rmd t} \right \rangle_t 
-
\left \langle \frac{a}{\Delta} {\hat P} \right \rangle_\lambda
\left \langle \frac{\rmd J_\phi}{\rmd t} \right \rangle_t
-
\Upsilon_{r}
\left \langle \frac{\rmd J_{r}}{\rmd t} \right \rangle_t\,;
\end{equation}
the factors $\langle {(r^2 + a^2){\hat P}}/{\Delta} \rangle_\lambda
,\, \langle {a {\hat P}}/{\Delta} \rangle_\lambda$ and $\Upsilon_{r}$, 
which are all local to the orbit, are defined in Sec.~\ref{subsec:dJdP}.  
Substituting the flux formulae of Eq.~\eqref{dJdt-nonres0} 
into these equations, the end results then reduce to those for 
$
\langle {\rmd {\hat E}}/{\rmd t} \rangle_t,\, 
\langle {\rmd {\hat L}}/{\rmd t} \rangle_t 
$, and $\langle {\rmd {\hat Q}}/{\rmd t} \rangle_t$
in the non-resonant case 
that were first presented by Sago et al.~\cite{Sago:2005gd,Sago:2005fn}.  
When instead substituting Eqs.~\eqref{dJdt-res0}, 
we arrive at precisely the same expressions 
for $\langle {\rmd {\hat E}}/{\rmd t} \rangle_t$ and 
$\langle {\rmd {\hat L}}/{\rmd t} \rangle_t$ in the resonant case 
that were first reported by Grossman, Levin, and Perez-Giz~\cite{Grossman:2011im} and Flanagan, Hughes, and Ruangsri~\cite{Flanagan:2012kg}. 
However, we find that our on-resonance expression 
for $\langle {\rmd {\hat Q}}/{\rmd t} \rangle_t$ 
only partially agrees with that presented in Ref.~\cite{Flanagan:2012kg}; 
the difference is attributed to the fact that 
Ref.~\cite{Flanagan:2012kg} only considers the radiative sector 
of Eq.~\eqref{dotQ-ave0} obtained from $\Hrad$, 
and discards all contributions to $\langle {\rmd J_{r}}/{\rmd t} \rangle_t$ 
coming from $\Hsym$ by construction (see Eqs.~\eqref{Flux-res}).~\footnote{The conservative self-force contribution 
to $\langle {\rmd {\hat Q}}/{\rmd t} \rangle_t$ 
in the resonant case may be explained in a heuristic way, 
without invoking technical Hamiltonian analysis. 
Consider a purely conservative self-forced dynamics in the corotating frame 
with the angular velocity of the periastron advance;  
the orbital plane does not precess in this frame. 
Introduce the axes $({{\tilde x}},\,{{\tilde z}})$ 
in the corotating frame to point the direction connecting 
the center of the background Kerr black hole and the periastron, 
and that normal to the orbital plane, respectively. 
In the non-resonant case, the orbit can ergodically fill up 
the orbital plane (bounded by two radii 
$r_{\rm min} \leq r \leq r_{\rm max}$) 
that has a ${\tilde x}$-${\tilde z}$ plane symmetry, 
and so there should be no ``torque of the conservative self-force'' 
in the ${\tilde x}$ direction (in a certain time-averaged sense). 
In contrast, the resonant orbit with a different 
initial value of $w^{\alpha}_0$ traces out a different ``shape'' 
in the orbital plane and breaks the plane symmetry, giving rise 
to the ``torque'' in the ${\tilde x}$ direction. 
Because this torque induces the change in the orbital inclination angle 
related to the Carter constant $Q$ (see, e.g., Refs.~\cite{Drasco:2005kz,Ganz:2007rf}), we can see that $\langle {\rmd {\hat Q}}/{\rmd t} \rangle_t$ on resonance would have the conservative contribution of the self-force 
(i.e., $\Hsym$).}

We conclude this introduction by briefly discussing 
some open issues and prospects for our results. 
First, it seems that the flux formulae 
of Eqs.~\eqref{dJdt-nonres0} and~\eqref{dJdt-res0} 
would suggest the ``balance laws'' for the actions $J_{\alpha}$ 
of the radiative inspiral of a small body into a Kerr black hole 
(in a time-averaged sense). 
At a conceptual level, our interpretation above sounds natural 
at least for $\langle {\rmd J_{t}}/{\rmd t} \rangle_t$ and 
$\langle {\rmd J_{\phi}}/{\rmd t} \rangle_t$
because the corresponding fluxes on their right-hand sides 
in Eqs.~\eqref{dJdt-nonres0} and~\eqref{dJdt-res0} coincide 
precisely with the fluxes of energy and axial angular momentum 
that gravitational radiation carries to infinity 
and down to the horizon~\cite{Sasaki:2003xr,Drasco:2005kz,
Grossman:2011im,Flanagan:2012kg}. 
However, the similar interpretation 
for $\langle {\rmd J_{r}}/{\rmd t} \rangle_t$  
and $\langle {\rmd J_{\theta}}/{\rmd t} \rangle_t$ is subtle. 
While the forms of Eqs.~\eqref{dJdt-nonres0} and~\eqref{dJdt-res0} would 
resemble more closely the ``balance laws'' than that of 
Eq.~\eqref{dotQ-ave0} for the Carter constant, 
still, the precise physical meaning of the ``out/down'' 
gravitational-wave fluxes encoded in 
$\langle {\rmd J_{r}}/{\rmd t} \rangle_t$  
and $\langle {\rmd J_{\theta}}/{\rmd t} \rangle_t$ are not known to us. 
For example, how does the background Kerr geometry 
evolve due to the radiative losses of the ``fluxes'' 
of $J_r$ and $J_{\theta}$? This question remains an open issue, 
and perhaps the answer may be provided through the ongoing development 
of the second-order formalism for the perturbed Einstein field
equation~\cite{Pound:2015wva,Miller:2016hjv}.

Second, the flux formulae of Eqs.~\eqref{dJdt-nonres0} and~\eqref{dJdt-res0} 
can be straightforwardly implemented, 
making use of the analytical/numerical Teukolsky-based flux codes 
developed by ``B.H.P.C.''~\cite{Fujita:2004rb,Sago:2005fn,
Fujita:2009us,Ganz:2007rf,
Fujita:2014eta,Sago:2015rpa,BHPC} with some mild adaptations 
(or any of the existing Teukolsky platforms such as the ones of 
Refs.~\cite{Drasco:2005kz,Sundararajan:2008zm,Shah:2014tka,Harms:2014dqa,Flanagan:2012kg,vandeMeent:2017bcc,BHPT}).
Because the flux formulae are built on the radiative sector of 
the averaged Hamilton equation of Eq.~\eqref{Jdot0}, 
it is completely equivalent to the effect of standard time-averaged 
dissipative self-force for generic (resonant) orbits in the Kerr geometry. 
Furthermore, the flux formulae are a (quasi-)gauge-invariant 
characterization  of the radiative dynamics for such generic orbits. 
Therefore, we expect that concrete calculation of the flux formulae 
will clarify the interesting possibility 
of ``sustained'' resonance in extreme mass-ratio inspirals 
(the orbit ``stuck'' on the resonance)~\cite{Isoyama:2013yor,vandeMeent:2013sza}, 
help in refining known practical schemes for simulating radiative 
extreme-mass-ratio inspirals~\cite{Mino:2005an,Tanaka:2005ue,Pound:2007th,Gair:2010iv,vandeMeent:2018rms}, 
and provide (yet another) accurate strong-field benchmark 
for the extreme mass-ratio regime of a more generic two-body problem 
in general relativity (see, e.g., Ref.~\cite{Tiec:2014lba}). 
We shall leave the task of actually producing  
the numerical data or analytical approximations 
to our forthcoming publication.

\section{Hamiltonian dynamics in the perturbed Kerr geometry with radiation}
\label{sec:Hamilton}

To set the stage, we begin with an extension of the Hamiltonian formulation 
of conservative dynamics in the perturbed Kerr geometry~\cite{Fujita:2016igj} 
to incorporate the full, physical, retarded metric perturbation sourced 
by a point particle. 
For the most part we shall import many helpful results from Sec.~II 
of Ref.~\cite{Fujita:2016igj}, but the reader should bear in mind that 
the discussion presented here allows for gravitation radiation 
from the particle. 
Our development below is limited to the self-force theory 
at first order in the mass ratio for the sub-extremal Kerr spacetime:  
we shall not be concerned with the second-order perturbation
theory~\cite{Harte:2011ku,Gralla:2012db,Pound:2012nt,Pound:2015wva} 
and the (perturbed) motion in the extremal Kerr geometry.

Throughout this work the Kerr metric $g_{\mu \nu}^{(0)}$ 
of mass $M$ and spin $S \equiv aM (< M^2)$ will be written
in terms of Boyer-Lindquist coordinates $(t,r,\theta,\phi)$. 
It is given by 
\begin{align}\label{Kerr} 
 g_{\mu \nu}^{(0)} \, \rmd x^{\mu} \rmd x^{\nu} = 
 &- \left( 1 - \frac{2 M r}{\Sigma} \right) \rmd t^2 
 - \frac{4 M a r \, {\sin^2 \theta} }{\Sigma} \, \rmd t \, \rmd \phi 
 + \frac{\Sigma}{\Delta} \, \rmd r^2 \nonumber \\ &+ \Sigma \, \rmd \theta^2 
 + \left( r^2 + a^2 + \frac{2 M a^2 r}{\Sigma} \sin^2 \theta \right)
 \sin^2 \theta \, \rmd \phi^2 \,,
\end{align}
where
\begin{equation}\label{Kerr-hojo}
\Sigma \equiv r^2 + a^2 \cos^2 \theta \,, 
\quad
\Delta \equiv r^2 - 2 M r + a^2 \,. 
\end{equation}
The mass of the particle is $\mu$ ($\ll M$) and 
the mass ratio is defined by $\eta \equiv \mu/M$. 
We will assume that particles' orbits in the limit $\eta \to 0$ recover 
generic bound geodesic orbits in Kerr geometry, 
restricted in $r_{\rm min} \leq r \leq r_{\rm max}$ and 
$\theta_{\rm min} \leq \theta \leq \pi - \theta_{\rm min}$, respectively.

\subsection{$4$D geodesic Hamiltonian and Hamilton's equations}
\label{subsec:Heq}

The Hamiltonian formulation in Ref.~\cite{Fujita:2016igj} 
begins with the observation that motion of a small particle is 
the geodesic motion in the smooth vacuum effective
metric~\cite{Detweiler:2002mi,Pound:2017psq}: 
\begin{equation}\label{g-eff}
g_{\mu \nu}(x; \gamma) \equiv 
g_{\mu \nu}^{(0)} (x) + h_{\mu \nu}^{(\text{R})}(x; \gamma) + O(\eta^2)\,,  
\end{equation}
where $g_{\mu \nu}^{(0)}$ is the background Kerr geometry, 
and $h_{\mu \nu}^{(\text{R})} = O(\eta)$ 
is the regular-part of the metric perturbation ($\equiv$ the R field) 
that is defined by subtracting an appropriate
singular-part metric perturbation 
($\equiv$ the S field) $h_{\mu \nu}^{(\text{S})}(x; \gamma)$
from a physical, retarded metric perturbation 
$h_{\mu \nu}^{+}(x; \gamma)$ generated 
by the source orbit $\gamma$~\cite{Pound:2015fma}.
At first order in the mass ratio, the R field can always be decomposed 
into the time-antisymmetric (``radiative'') field 
$h_{\mu \nu}^{(\text{rad})} 
\equiv \frac{1}{2} \, (h_{\mu\nu}^{+} - h_{\mu\nu}^{-})$ 
and the (regularized) time-symmetric field 
$h_{\mu \nu}^{(\text{sym})} 
\equiv \frac{1}{2} \, (h_{\mu\nu}^{+} + h_{\mu\nu}^{-})
- h_{\mu \nu}^{(\text{S})}$ with the advanced metric perturbation 
$h_{\mu \nu}^{-}(x; \gamma)$, and Ref.~\cite{Fujita:2016igj} 
discussed only the symmetric field 
instead of the full R field to define conservative dynamics.  
Despite this difference in the metric perturbation, 
the essential picture remains unchanged: 
the particle moves on a geodesic of the effective metric. 
The generalization of the Hamiltonian formulation~\cite{Fujita:2016igj} 
to the full effective metric $g_{\mu \nu}$ of Eq.~\eqref{g-eff} is therefore 
immediate. 

In our framework, the geodesic Hamiltonian in the effective metric 
is a standard $4$D Hamiltonian. We write this as 
\begin{equation}\label{defH}
H(x,u;\gamma) \equiv 
\frac{1}{2} \, g^{\mu \nu}(x;\gamma) \,u_{\mu} u_{\nu} \,,
\end{equation}
or, equivalently, expand it as 
\begin{equation}\label{H0H1}
H(x,u;\gamma)= H^{(0)} (x, u) + \Hint (x, u; \gamma) + O(\eta^2)\,.
\end{equation}
where $H^{(0)}(x,u)$ is the unperturbed background Hamiltonian,
simply defined by the expression in Eq.~\eqref{defH} 
with the substitution $g^{\mu\nu} \to g_{(0)}^{\mu\nu}$, and 
\begin{equation}\label{Hint}
\Hint(x, u; \gamma)
\equiv 
- \frac{1}{2} {h}^{\mu \nu}_{(\text{R})} (x; \gamma) \, u_{\mu} u_{\nu} \,, 
\end{equation}
is \textit{the perturbed interaction Hamiltonian} ($\propto\eta$). 
There is no need to display a more explicit form 
of $h_{\mu \nu}^{(\text{R})}$ in this section, 
and we shall defer it to Sec.~\ref{sec:rad-sym}. 

The Hamiltonian of Eq.~\eqref{defH} leads to Hamilton equations 
for the canonical position $x^\mu$ and momentum $u_\mu$: 
\begin{equation}\label{H-eq0}
 \dot x^{\nu} = 
 \left(\frac{\partial H}{\partial u_{\nu}}\right)_{\!x} \,,
 \quad
 \dot u_{\nu} = - \left(\frac{\partial H}{\partial x^{\nu}}\right)_{\!u} \,,
\end{equation}
where the overdot stands for the derivative 
with respect to the proper time $\tau$ along the orbit. 
Here, it is important to recognize that $\tau$ is measured 
in the effective metric of Eq.~\eqref{g-eff}, 
not in the background Kerr metric $g_{\mu \nu}^{(0)}$.
As a consequence, the canonical variables $(x^{\mu},\,u_{\mu})$ 
have to be normalized according to 
\begin{equation}\label{norm-xu}
\left. g^{\mu \nu} u_{\mu} u_{\nu}\right|_\gamma  
= 
\left. {\dot x}^{\mu} u_{\mu}\right|_\gamma  
=- 1 \,, 
\end{equation}
for any physical orbits (i.e., on-shell solutions of Eqs.~\eqref{H-eq0}).

\subsection{Constants of motion and action-angle variables}
\label{subsec:wJ}

Next, we wish to introduce generalized action-angle variables 
$(w^{\alpha},J_{\alpha})$~\cite{Schmidt:2002qk,Hinderer:2008dm}, 
which result from the two-step canonical transformations 
$(x^{\mu},\, u_{\mu}) \leftrightarrow (X^{\alpha},\, P_{\alpha})
\leftrightarrow (w^{\alpha},\, J_{\alpha})$ described below. 

For these canonical transformations, 
we first need a set of canonical variables $(X^{\alpha},\, P_{\alpha})$ 
such that the canonical momenta $P_\alpha$ recover the constants of motion 
for Kerr geodesics in the test-mass limit $\eta \to 0$~\cite{Schmidt:2002qk}. 
Thanks to the symmetry of Kerr geometry associated with 
the Killing vectors $t^{\mu}$ and $\phi^{\mu}$,  
and the Killing tensor $K^{\mu \nu}$~\cite{Walker:1970un} 
(see Appendix~\ref{sec:Killing} for their explicit expressions), 
the Kerr geodesics admit the three non-trivial constants of motion: 
the specific energy $\hat E$,  azimuthal angular momentum $\hat L$, 
and the \textit{Carter constant} ${\hat Q}$~\cite{Carter:1968rr}. 
In the Hamiltonian formalism of Ref.~\cite{Fujita:2016igj}, 
$\hat E$, $\hat L$, and ${\hat Q}$ are all promoted 
to canonical momenta $P_{\alpha}$ on the phase space to define 
\begin{equation}\label{def-P}
 P_0 = -\frac{\hat\mu^2}{2} 
 \equiv H^{(0)} \,,
 \quad
 P_1 = \hat E \equiv - t^{\mu} u_{\mu} \,, 
 \quad
 P_2 = \hat L \equiv  \phi^{\mu} u_{\mu}\,, 
 \quad
 P_3 = {\hat Q} \equiv K^{\mu \nu} u_{\mu} u_{\nu}\,,
\end{equation}
where we introduce ${\hat \mu}$ to complete the $4$D canonical momenta:    
it is important to distinguish the canonical variable 
${\hat {\mu}}$ from the physical mass of the particle $\mu$.
Explicitly, ${\hat E} \equiv - u_{t}$, ${\hat L} \equiv u_{\phi}$, and 
\begin{equation}\label{Q}
{\hat Q} 
=
a^2 \cos^2 \theta ({\hat \mu}^2 - {\hat E}^2) 
+ \cot^2 \theta {\hat L}^2 + u_\theta^2 
+ ( a {\hat E} - {\hat L})^2 
=
\frac{{\hat P}^2}{\Delta} - \Delta\, u_r^2 - {\hat \mu}^2 r^2  \,,
\end{equation}
where ${\hat P} \equiv {\hat E} (r^2 + a^2) -  a {\hat L}$,
and the second equality follows from the ``duality'' 
of $K_{\mu \nu}$ (see Eq.~\eqref{K-rth}). 
From Eqs.~\eqref{def-P}, the first transformation 
between $(x^{\mu},\, u_{\mu})$ and $(X^{\alpha},\, P_{\alpha})$ can be defined 
by the (type II) generating function~\cite{Carter:1968rr},
\begin{equation}\label{def-W}
 W(x,\, P) = -{\hat E}\,t + {\hat L} \, \phi
 + \int^r \frac{\sqrt{R (r',\, P)}}{\Delta} \, \rmd r' 
 + \int^{\theta} {\sqrt{\Theta (\cos \theta',\,P)}}\,\rmd \theta' \,,
\end{equation}
with~\footnote{
The functions $\sqrt{R (r,\, P)}$ and $\sqrt{\Theta (\cos \theta,\, P)}$ 
of Eqs.~\eqref{def-W} are positive (negative) 
when $\dot r$ and $\dot \theta$ are positive (negative).
}
\begin{subequations}\label{def-R-Theta}
 \begin{align}
R(r,\,P) &\equiv 
{\hat P}^2 - \Delta \left( {\hat Q} + {\hat \mu}^2 r^2  \right)\,, \\
\Theta (\cos \theta,\, P) &\equiv 
{\hat C} - \left\{ \left({\hat \mu}^2 - {\hat E}^2\right) a^2 
+ \frac{ {\hat L}^2}{1 - \cos^2 \theta} \right\} \cos^2 \theta \,,
 \end{align}
\end{subequations}
where another canonical variable related to the Carter constant, 
\begin{equation}\label{C}
{\hat C} \equiv {\hat Q} - (a {\hat E} - {\hat L})^2 
=
a^2 ({\hat \mu}^2 - {\hat E}^2) \cos^2 \theta 
+ \cot^2 \theta {\hat L}^2 + u_\theta^2\,, 
\end{equation}
is introduced for future reference; 
notice that ${\hat Q} = {\hat Q}(r,\, u)$ 
while ${\hat C} = {\hat C}(\cos \theta,\,u)$~\footnote{
Note that there are some differences in the notation 
of the ``Carter constants'' ${\hat Q}$ and ${\hat C}$, 
which might bring confusion: 
for historical reasons, much of the Kyoto literature uses 
the notation of Eqs.~\eqref{Q} and~\eqref{C}, while some literature 
(e.g., the North American works~\cite{Hinderer:2008dm,
Flanagan:2012kg}) may use different symbols, namely, 
${\hat Q}^{\text{Kyoto}} = {\hat K}^{\text{NA}}$ 
and ${\hat C}^{\text{Kyoto}} = {\hat Q}^{\text{NA}}$, respectively; 
see also the chapter 33 of Ref.~\cite{Misner:1974qy}}.

From Eq.~\eqref{def-W}, we obtain 
\begin{equation}\label{def-p0}
 u_t  = - P_1 \,, 
 \quad
 u_{\phi} = P_2 \,, 
 \quad
 u_r = \frac{\sqrt{R(r,\,P)}}{\Delta} \,, 
 \quad
 u_{\theta} = \sqrt{{\Theta}(\cos \theta,\,P)} \,,
\end{equation} 
making use of $ u_{\mu} = ({\partial W}/{\partial x^{\mu}} )_{P}$. 
Explicit expressions for $X^{\alpha}(x,\,u)$ will not be needed in this paper; 
they are displayed in Eq.~(2.17) of Ref.~\cite{Fujita:2016igj}.

We next use Eqs.~\eqref{def-p0} to introduce action variables $J_{\alpha}$. 
They are defined by~\cite{Schmidt:2002qk,Hinderer:2008dm}
\begin{equation}\label{def-J0}
 J_t \equiv - P_1 \,, 
 \quad
 J_{\phi} \equiv P_2 \,, 
 \quad
 J_{r} \equiv 
 \frac{1}{2 \pi} \oint \frac{\sqrt{R(r,\,P)}}{\Delta} \, \rmd r \,, 
 \quad
 J_{\theta} \equiv 
 \frac{1}{2 \pi} \oint {\sqrt{\Theta (\cos \theta,\,P)}} \, \rmd \theta \,,
\end{equation}
where $\oint$ denotes twice the integral over the allowed region of motion 
described by $R(r,\,P) \geq 0$ and $\Theta(\cos \theta,\,P) \geq 0$. 
By definition, we have 
\begin{equation}\label{JvsP}
 J_{\alpha} = J_{\alpha}(P)\,,
\end{equation} 
and $J_{\alpha}(P)$ is an invertible function to give $P_{\mu} = P_{\mu}(J)$, 
as shown in Ref.~\cite{Schmidt:2002qk}. 
This allows us to rewrite the generating function 
of Eq.~\eqref{def-W} as ${\cal W}(x,\, J) \equiv W(x,\, P(J))$, 
and it then generates the desired canonical transformation 
between $(x^{\mu},\, u_{\mu})$ and the action-angle variables 
$(w^{\alpha},\, J_{\alpha})$. 
The results are 
$u_{\mu} = ({\partial {W}}/{\partial x^{\mu}} )_{J}$ and 
\begin{equation}\label{def-w}
w^{\alpha} 
= 
{\breve w}^{\alpha} + 
\biggl(\frac{\partial {\cal W}}{\partial J_{\alpha}}\biggr)_{\!r,\theta}\,,
\end{equation}
where 
${\breve w}^{\alpha} \equiv \{t,\, 2 \pi N^r,\, 2 \pi N^\theta ,\,\phi \}$ 
with integers $N^{r}$ and $N^{\theta}$, and 
${\cal W}(r,\, \theta,\, J) \equiv 
\int_{r_{\mathrm {min}}}^r \!\! (\sqrt{{R}}/{\Delta}) \, \rmd r' 
+ 
\int_{\theta_{\mathrm {min}}}^{\theta} \!\! \sqrt{\Theta} \, \rmd \theta'$. 

With these action-angle variables, Hamilton's canonical equations read 
\begin{equation}\label{Heq-wJ}
{\dot w}^{\alpha}
= \left(\frac{\partial H}{\partial J_{\alpha}}\right)_{\!w} \,, 
\quad
\dot J_{\alpha} 
= 
- \left(\frac{\partial H}{\partial w^{\alpha}}\right)_{\!J} 
=
- \left(\frac{\partial \Hint}{\partial w^{\alpha}}\right)_{\!J} \,, 
\end{equation}
where we use the fact that $(\partial H^{(0)}(J) / \partial w^{\alpha})_J = 0$ 
in the second equality of the latter equation: 
recall Eqs.~\eqref{def-P}. 
It is important to recognize that $(w^{\alpha},\, J_{\alpha})$ 
evaluated along the physical orbit $\gamma$ must satisfy 
\begin{equation}\label{norm-wJ}
\left.{\dot w^\alpha} J_\alpha \right|_{\gamma}
= -1\,
\end{equation}
due to the normalization condition of Eq.~\eqref{norm-xu}. 
While this relation seems to have been unnoticed in many papers, 
it is now recognized as an algebraic formula to be interpreted as 
the ``first law of binary mechanics''
(in the self-force theory)~\cite{Tiec:2013kua,Fujita:2016igj},~\footnote{
``The first law of binary mechanics'' is a conjectured variational formula 
for (point-particle) binary systems that relates 
the local property of each body to either global conserved charges 
of the binary system (the Bondi binding energy, etc.) 
or local conserved quantities along the orbit   
(the mechanical energy ${\hat E}$, etc). 
The first law has been formulated in a number of context 
of two-body dynamics in general relativity, 
including the post-Newtonian method, self-force theory 
and numerical-relativity simulation; the recent developments are reviewed 
in Sec.~3.2.4, Chapter II, of Ref.~\cite{Barack:2018yly}.}
and plays a crucial role in reading out the physical effects 
of the self-force with the Hamiltonian formalisms~\cite{Isoyama:2014mja}. 

\subsection{Orbital resonance in Kerr geodesics}
\label{subsec:def-res}

The manipulation of the Hamilton equations of Eq.~\eqref{Heq-wJ} 
will require as input the (source) orbit $\gamma$ 
for the interaction Hamiltonian $\Hint(w,\,J; \gamma)$, 
and it is approximated by generic bound geodesic orbits 
in background Kerr spacetime.
For future reference, we now 
take the test-mass limit $\eta \to 0$ in Eq.~\eqref{Heq-wJ} 
and obtain the solutions 
$(w^{\alpha}(\tau),\, J_{\alpha}(\tau))$ for the bound Kerr geodesics.

The spatial components of the angle variables 
$\{w^{r},\,w^{\theta},\,w^{\phi}\}$ given 
by Eqs.~\eqref{def-w} are $2 \pi$-periodic coordinates on the phase space. 
This observation then implies that the Hamilton's equation 
for ${\dot w^{\alpha}}$ can define the angular frequencies of Kerr geodesics 
with respect to the proper time $\tau$~\cite{Schmidt:2002qk,Hinderer:2008dm,Tiec:2013kua,Fujita:2016igj},
\begin{equation}\label{def-omega}
\omega^{\alpha} 
\equiv
{\dot w}^{\alpha} 
=
\left( z^{-1},\,\omega^r,\,\omega^\theta,\,\omega^\phi \right)\,,
\quad 
\Omega^{\alpha} \equiv z\,\omega^{\alpha}\,,
\end{equation}
where we have introduced the (background) ``redshift'' variable  
$z \equiv (\omega^{t})^{-1}$ in the phase space
and the associated fundamental frequencies $\Omega^{\alpha}$ measured 
with respect to the Boyer-Lindquist coordinate time $t$   
(i.e., the proper time of a static observer in the asymptotically
far region)~\footnote{
We should mention the fact that the fundamental frequencies 
$\Omega^{\alpha}$ can be ``degenerate'' in the sense that 
a pair of physically distinct Kerr geodesics can share the same 
values of $\Omega^{\alpha}$. 
Such pairs of Kerr geodesics are now known as
\textit{isofrequency pairs}~\cite{Barack:2011ed,Warburton:2013yj}, 
but this arises no difficulties in our analysis.
}.
With Eq.~\eqref{def-omega}, the solutions to the Hamilton equations 
of Eqs.~\eqref{Heq-wJ} can be expressed as 
\begin{equation}\label{soln-w}
w^{\alpha} (\tau) = \omega^{\alpha}\,(\tau - \tau_0) + w^{\alpha}_0 \,, 
\end{equation}
where $\tau_0$ and $w^{\alpha}_0$ are some initial values, 
and the constants of motion $J_{\alpha}(\tau) = J_{\alpha}$ 
given by Eqs.~\eqref{def-J0}.  

When the radial and angular motions become commensurable, 
there exists only one fundamental frequency $\tilde{\Omega}$ 
for their motions.
This is the $r\mathchar`-\theta$ orbital resonance, 
which will require a separate treatment in our analysis.
For the fundamental frequencies $\Omega^r$ and $\Omega^{\theta}$ 
of Eqs.~\eqref{def-omega}, we define the resonant relation 
for Kerr geodesics by 
\begin{equation}\label{res-Omega}
\tilde{\Omega} \equiv 
\frac{\Omega^r}{\beta^r} 
= 
\frac{\Omega^{\theta}}{\beta^{\theta}} \,, 
\end{equation}
with a pair of coprime integers $\{\beta^r,\,\beta^\theta\}$, 
and the associated resonant phase by 
\begin{equation}\label{D-perp}
w^{\perp}_0 
\equiv 
\frac{w^{\theta}_0}{\beta^{\theta}}
-
\frac{w^{r}_0}{\beta^{r}}\,.
\end{equation}

The initial values $w^{\alpha}_0$ in Eq.~\eqref{soln-w} 
will play a very important role in the next two sections.  
Not all of them, however, are relevant to our discussion. 
Without loss of generality, we have 
\begin{equation}\label{w0-tphi}
{w^{t}_0} = 0 = {w^{\phi}_0}\,,
\end{equation}
whether or not the orbit is resonant 
because the Kerr geometry is stationary and axially symmetric:
recall Eqs.~\eqref{def-w}. 
While $w^{r}_0$ and $w^{\theta}_0$ cannot be set 
to zero making use of the symmetry of the Kerr geometry in general, 
we are still allowed to have 
\begin{equation}\label{w0-rth}
{w^{r}_0} = 0 \quad \text{or} \quad {w^{\theta}_0} = 0\,
\end{equation}
modulo $2\pi$ by selecting a suitable value of $\tau_0$. 
If the orbit is non-resonant, Eq.~\eqref{w0-rth} further yields 
${w^{r}_0} = 0 = {w^{\theta}_0}$ thanks to the fact that 
the orbital periods $2\pi / \omega^{r}$ and $2\pi / \omega^{\theta}$ 
are incommensurable; this is essentially the same argument 
as first produced by Mino~\cite{Mino:2003yg}. 
The above dependence of the initial phase of the resonant orbit
was stressed in a number of works~\cite{Grossman:2011im,Flanagan:2012kg,
Isoyama:2013yor,vandeMeent:2013sza,Brink:2015roa}, 
and our observation here agrees with their analysis.

\section{Green's-function-based definition of the interaction Hamiltonian}
\label{sec:rad-sym}

In this section, we simplify the Hamilton equations 
of ${\dot J_{\alpha}}$ in Eqs.~\eqref{Heq-wJ} 
by taking a long-time average, defining 
\begin{equation}\label{longtime}
\langle f \rangle_{s} 
\equiv 
\lim_{T\to\infty}\frac{1}{2T}\int_{-T}^{T} {f}(s) \,\rmd s \,, 
\end{equation} 
for various functions $f(s) = O(\eta)$ along 
the \textit{background} orbits parameterized by $s$ 
(e.g., the coordinate time $t$ or proper time $\tau$). 
This is essentially equivalent to working in the leading-order  
two-timescale (``adiabatic'') approximation of Eqs.~\eqref{Heq-wJ}---
see, e.g., Refs.~\cite{Pound:2007ti,Hinderer:2008dm,Flanagan:2010cd}.~\footnote{It should be noted that the long-time average 
of Eq.~\eqref{res-dHsym} used for our analysis is not always equivalent 
to the phase-space average over the angle variables $w^{\alpha}$ 
in the literature 
(see, e.g., Ref.~\cite{Hinderer:2008dm} for the precise definition). 
The two averaging procedures are reconciled only when 
the orbits are non-resonant~\cite{Grossman:2011im}, 
and we need to be mindful of their difference in the resonant case.} 
We begin with the Green's-function-based expression 
for the interaction Hamiltonian of Eq.~\eqref{Hint} 
to split it into radiative and symmetric portions, 
and then summarize some of the pertinent properties 
of the averaged Hamilton equations based on that split; 
they are extensively studied in Sec. III and IV 
of Ref.~\cite{Fujita:2016igj}, 
to which the discussion here refers the reader for details.

\subsection{Interaction Hamiltonian: the radiative-symmetric decomposition}
\label{subsec:Hint-rad}

A key property of the interaction Hamiltonian of Eq.~\eqref{Hint} is 
the functional dependence of the source orbit $\gamma$ with which 
the R-field $h_{\mu \nu}^{(\text{R})}(x; \gamma)$
is generated~\cite{Pound:2015tma}. 
This aspect becomes especially clear in view of the Green's-function-based 
definition, which is written as
\begin{align}\label{Hint-G}
 \Hint(x,u;\gamma)
 =
 - \frac{1}{2} {h}^{\mu \nu}_{(\text{R})} (x; \gamma) \, u_{\mu} u_{\nu}
 =
 -\frac{\mu}{2}
 \left.
 \int \rmd \tau' \,
 u_{\mu} u_{\nu}\, 
 G^{\,\mu\nu\,\rho\sigma}_{\mathrm {(R)}} \left(x; x'\right)
 \,u'_\rho u'_\sigma
 \right|_{\substack{x'=x(\tau') \\ u'=u(\tau')}} \,,
\end{align}
where
$G^{\,\mu\nu\,\rho\sigma}_{\mathrm {(R)}}(x; x')
\equiv 
G^{\mu\nu\,\rho\sigma}_{+}(x; x')
-
G^{\mu\nu\,\rho\sigma}_{\mathrm {(S)}}(x; x')$ is the R-part 
of the Green's function for the linearized Einstein equation 
defined by the retarded Green's function 
$G^{\mu\nu\,\rho\sigma}_{+}(x; x')$ and the S-part of the Green's function 
$G^{\mu\nu\,\rho\sigma}_{\mathrm {(S)}}(x; x')$~\cite{Detweiler:2002mi}; 
the details are reviewed in Sec.~16 of Ref.~\cite{Poisson:2011nh}. 
The variables with a prime here correspond 
to the (source) orbit $\gamma$ approximated by the generic bound geodesics 
in Kerr spacetime (at leading order in the two-timescale approximation).

We next turn to decompose $\Hint$ into the antisymmetric 
(``radiative'') pieces $\Hrad$ and ``symmetric'' piece $\Hsym$, 
adopting the split of $G^{\mu\nu\,\rho\sigma}_{\mathrm {(R)}}$ 
into antisymmetric and symmetric portions~\footnote{
We have omitted the superscript ``(1)'' from $\Hrad$ and $\Hsym$
as we shall focus only on the first-order effects.
}.
Following Gal'tsov~\cite{Galtsov:1982hwm} and Mino~\cite{Mino:2003yg}, 
we define antisymmetric radiative Green's function 
$G^{\mu\nu\,\rho\sigma}_{\mathrm {(rad)}}$ 
and the (regularized) symmetric Green's function 
$G^{\mu\nu\,\rho\sigma}_{\mathrm {(sym-S)}}$ by  
\begin{subequations}\label{radsym-G}
\begin{align}
G^{\mu\nu\,\rho\sigma}_{\mathrm {(rad)}}(x; x')
&\equiv
\frac{1}{2}
\left\{
G^{\mu\nu\,\rho\sigma}_{+} (x; x') 
-  
G^{\mu\nu\,\rho\sigma}_{-} (x; x')
\right\}\,, \\
G^{\mu\nu\,\rho\sigma}_{\mathrm {(sym-S)}}(x; x')
&\equiv
\frac{1}{2}
\left\{
G^{\mu\nu\,\rho\sigma}_{+}(x; x') 
+ 
G^{\mu\nu\,\rho\sigma}_{-}(x; x')
\right\}
-
G^{\mu\nu\,\rho\sigma}_{\mathrm {(S)}}(x; x')\,, 
\end{align}
\end{subequations}
where we have introduced the advanced Green's function
$G^{\mu\nu\,\rho\sigma}_{-}(x; x') 
= 
G^{\mu\nu\,\rho\sigma}_{+}(x'; x)$. 
Notice that $G^{\mu\nu\,\rho\sigma}_{\mathrm {(rad)}}(x; x')
= -G^{\mu\nu\,\rho\sigma}_{\mathrm {(rad)}}(x'; x)$, 
and 
$G^{\mu\nu\,\rho\sigma}_{\rm (sym-S)}(x; x')
=
G^{\rho\sigma\,\mu\nu}_{\rm (sym-S)}(x'; x)$.
Armed with such definitions, 
we simply define $\Hrad$ and $\Hsym$ by the substitutions of 
$G^{\,\mu\nu\,\rho\sigma}_{\mathrm {(R)}}
\to 
G^{\,\mu\nu\,\rho\sigma}_{\mathrm {(rad)}}$ 
and 
$G^{\,\mu\nu\,\rho\sigma}_{\mathrm {(R)}}
\to 
G^{\,\mu\nu\,\rho\sigma}_{\mathrm {(sym-S)}}$
in the expression of Eq.~\eqref{Hint-G}, respectively.

\subsection{Averaged Hamilton equation}
\label{subsec:ave-Heq}

The radiative-symmetric split of $\Hint$ is convenient 
for a number of reasons, and its most important advantage 
is that the $\tau$-averaged Hamilton's equation 
$\langle {\dot J_{\alpha}} \rangle_\tau$ is simplified to 
\begin{equation}\label{Jdot-nonres}
\langle \dot J_{\alpha} \rangle_\tau
= -\left\langle \left(
\frac{\partial \Hrad}{\partial w^{\alpha}} \right)_{\!J}
\right\rangle_\tau \,,
\end{equation}
for the non-resonant case, and
\begin{equation}\label{Jdot-res}
\langle \dot J_{\alpha} \rangle_\tau
= 
-\left\langle \left(
\frac{\partial \Hrad}{\partial w^{\alpha}} \right)_{\!J}
\right\rangle_\tau 
- (\delta_{\alpha}^{r} +  \delta_{\alpha}^{\theta} )
\left\langle \left(
\frac{\partial \Hsym}{\partial w^{\alpha}} \right)_{\!J}
\right\rangle_\tau \,,
\end{equation}
for the resonant case (recall Eq.~\eqref{res-Omega}).

The points behind the proof of Eqs.~\eqref{Jdot-nonres} and~\eqref{Jdot-res} 
are that (i) the symmetric interaction Hamiltonian $\Hsym$,
which is defined by Eq.~\eqref{Hint-G} with the substitution 
$G^{\,\mu\nu\,\rho\sigma}_{\mathrm {(R)}}
\to 
G^{\,\mu\nu\,\rho\sigma}_{\mathrm {(sym-S)}}$
is symmetric 
under the exchange of the ``field variables'' $(x^{\mu},\,u_{\mu})$ 
and ``source variables'' $(x'^{\mu},\,u'_{\mu})$,
and that
(ii) the $\tau$-averaged symmetric interaction Hamiltonian 
$\langle \Hsym \rangle_{\tau}$ depends only on the actions $J_{\alpha}$ 
and the initial values of the angle variables $w^{\alpha}_0$ 
in Eq.~\eqref{soln-w}. 
These two observations allow us to write  
\begin{equation}\label{del-Hsym}
\left\langle \left(
\frac{\partial \Hsym}{\partial w^{\alpha}} \right)_{\!J}
\right\rangle_\tau
=
\frac{1}{2} 
\frac{\delta }{\delta w^{\alpha}_0}
\left\langle \Hsym \right\rangle_\tau\,,
\end{equation}
where ${\delta}/{\delta w^{\alpha}_0}$ is the total variation 
with respect to $w^{\alpha}_0$, and we have used the identity 
\begin{align}\label{Jdot-id}
& \frac{1}{2} 
\frac{\delta }{\delta w^{\alpha}_0}
\left\langle \Hsym \right\rangle_\tau \cr 
& \quad =
\frac{1}{2}
\lim_{T \to \infty}
\frac{1}{2T} \int^{T}_{-T} \rmd \tau \int^{T}_{-T} \rmd \tau'
 \left.
 \left( \frac{\partial}{\partial w^{\alpha}} + \frac{\partial}{\partial w'^{\alpha}} \right)  G^{\mathrm {(sym)}}(x,\, u; x',\,u')
 \right|_{\substack{x=x(\tau), \, x'=x(\tau') \\ u=u(\tau), \, u'=u(\tau')}} 
 \cr 
& \quad =
\lim_{T \to \infty}
\frac{1}{2T} \int^{T}_{-T} \rmd \tau \int^{T}_{-T} \rmd \tau' \, 
 \left.
 \frac{\partial}{\partial w^{\alpha}} G^{\mathrm {(sym)}}(x,\,u; x',\,u')
 \right|_{\substack{x=x(\tau), \, x'=x(\tau') \\ u=u(\tau), \, u'=u(\tau')}} 
 \,,
\end{align}
with $G^{\mathrm {(sym)}}(x,\,u; x',\,u') \equiv
(-\mu/2) u_{\mu} u_{\nu}\, 
G^{\,\mu\nu\,\rho\sigma}_{\mathrm {(sym-S)}} (x; x')\,
u'_\rho u'_\sigma 
=
G^{\mathrm {(sym)}}(x',\,u'; x,\,u)$. 
The factor $1/2$ on the right-hand side of Eq.~\eqref{del-Hsym} accounts 
for ${\delta}/{\delta w^{\alpha}_0}$ acting on \textit{both} 
the field and source orbits 
parameterized by $w^{\alpha}(\tau)$ and $w'^{\alpha}(\tau)$, respectively,  
while the partial derivative on the left-hand side 
acts only on the field variable of $w^{\alpha}$.

Without loss of generality we can assume Eq.~\eqref{w0-tphi},
i.e., $\Hsym$ is independent of $w^{t}_0$ and $w^{\phi}_0$,
which implies 
\begin{equation}\label{dHsymdtphi}
\left\langle \left(
\frac{\partial \Hsym}{\partial w^{t}} \right)_{\!J}
\right\rangle_\tau 
= 0 =
\left\langle \left(
\frac{\partial \Hsym}{\partial w^{\phi}} \right)_{\!J}
\right\rangle_\tau\,,
\end{equation}
whether or not the orbits experience resonance. 
Similarly, one can further show that 
\begin{equation}
\left\langle \left(
\frac{\partial \Hsym}{\partial w^{r}} \right)_{\!J}
\right\rangle_\tau 
= 0 =
\left\langle \left(
\frac{\partial \Hsym}{\partial w^{\theta}} \right)_{\!J}
\right\rangle_\tau, 
\quad \text{for non-resonant orbits}\,,
\end{equation}
because we are always allowed to have $w^{r}_0 = 0 = w^{\theta}_0$, 
i.e., $\langle  \Hsym \rangle_\tau$  is independent 
of $w^{r}_0$ and $w^{\theta}_0$: 
recall the discussion in Sec.~\ref{subsec:def-res}. 
Equation~\eqref{Jdot-nonres} then follows.

For the resonant orbits, there is no known argument to guarantee 
$\langle ({\partial \Hsym}/{\partial w^{r}} )_{J} \rangle_\tau 
= 0 = 
\langle ({\partial \Hsym}/{\partial w^{\theta}} )_{J} \rangle_\tau$. 
It is possible, however, to make some progress 
by taking the linear combination 
$\beta^r 
\langle ({\partial \Hsym}/{\partial w^{r}} )_{J} \rangle_\tau 
+
\beta^{\theta} 
\langle ({\partial \Hsym}/{\partial w^{\theta}} )_{J} \rangle_\tau$ 
with the integers $\beta^{r}$ and $\beta^{\theta}$ of
Eq.~\eqref{res-Omega}, which characterize resonance. 
We recall that the $\tau$-derivative of $\Hsym = O(\eta)$ implies 
\begin{equation}\label{dHsymdtau}
{\dot H}_{\mathrm {sym}} 
=
{\dot w}^\alpha \left(
\frac{\partial \Hsym}{\partial w^{\alpha}} 
\right)_{\!J} 
+
{\dot J}_\alpha \left(
\frac{\partial \Hsym}{\partial J_{\alpha}} 
\right)_{\!w} 
=
{\omega}^\alpha \left(
\frac{\partial \Hsym}{\partial w^{\alpha}} 
\right)_{\!J} + O(\eta^2)\,,
\end{equation}
where we have used Hamilton's equation of Eq.~\eqref{Heq-wJ}. 
At the linear order in $\eta$, the $\tau$-average of the left-hand side of 
Eq.~\eqref{dHsymdtau} gives 
\begin{equation}\label{Hsym-ave}
\left \langle {\dot H}_{\mathrm {sym}} \right \rangle_\tau
=
\lim_{T\rightarrow\infty} 
\frac{1}{2T} \left[ {\Hsym (T) - \Hsym (-T)} \right] = 0\,
\end{equation}
because the orbit in the test-mass limit $\eta \to 0$ 
is the bound Kerr geodesics, and $\Hsym = O(\eta)$ does not 
secularly grow (in the class of gauge in which the effective metric 
of Eq.~\eqref{g-eff} is well defined~\cite{Pound:2015fma}). 
Inserting this and Eq.~\eqref{dHsymdtphi} into Eq.~\eqref{dHsymdtau} 
with Eq.~\eqref{res-Omega}, we arrive at 
\begin{equation}\label{res-Hsym}
\beta^r \left\langle \left(
\frac{\partial \Hsym}{\partial w^{r}} \right)_{\!J}
\right\rangle_\tau 
+ 
\beta^\theta \left\langle \left(
\frac{\partial \Hsym}{\partial w^{\theta}} \right)_{\!J}
\right\rangle_\tau 
= 0 \quad \text{for resonant orbits}\,.
\end{equation}
With Eq.~\eqref{del-Hsym} and the resonant phase 
$w^{\perp}_0$ of Eq.~\eqref{D-perp}, 
this relation can be translated to 
\begin{equation}\label{res-dHsym}
\frac{1}{2} 
\frac{\delta }{\delta w^{\perp}_0}
\left\langle \Hsym \right\rangle_\tau\, 
=
-\beta^r \left\langle \left(
\frac{\partial \Hsym}{\partial w^{r}} \right)_{\!J}
\right\rangle_\tau 
=
\beta^\theta \left\langle \left(
\frac{\partial \Hsym}{\partial w^{\theta}} \right)_{\!J}
\right\rangle_\tau \,.
\end{equation}
Equation \eqref{res-dHsym} confirms that the presence of 
$\langle ({\partial \Hsym}/{\partial w^{r}} )_{J} \rangle_\tau$
and 
$\langle ({\partial \Hsym}/{\partial w^{\theta}} )_{J} \rangle_\tau$ 
is a consequence of the initial-phase dependence of the resonant orbit. 
Thus, the general expressions for $\langle {\dot J_{\alpha}} \rangle_\tau$ 
should be given by Eq.~\eqref{Jdot-res}.~\footnote{
If we further average Eq.~\eqref{res-dHsym} with respect to $w^{\perp}_0$, 
which is essentially equivalent to the phase space average of $\Hsym$ 
over the $2$-torus parameterized by $w^{r}$ and $w^{\theta}$, 
it identically vanishes. This agrees with the conclusion 
of Ref.~\cite{Hinderer:2008dm} 
(i.e., the phase-space averaged rates of change of $J_{\alpha}$ 
do not have a contribution from the conservative self-force 
whether or not the orbit experiences resonance).}

\subsection{Quasi gauge invariance of $\langle {\dot J}_{\alpha} \rangle_\tau$}
\label{subsec:gauge}

Before proceeding, we remind the reader that our Hamiltonian formalism 
assumes on a certain restricted class of gauges 
in which the R-field perturbation $h_{\mu \nu}^{(\text{R})}$ 
is well defined everywhere around the orbit~\cite{Pound:2015fma}. 
Although Eqs.~\eqref{Heq-wJ} are altered by a gauge transformation,  
$\langle {\dot J_{\alpha}} \rangle_\tau$ 
are \textit{quasi gauge invariant} within that class of gauges.~\footnote{
The quantity $\langle {\dot J_{\alpha}} \rangle_\tau$ is only locally 
defined along the orbit, and is not gauge invariant 
in the strict mathematical sense in general relativity. 
In fact, the gauge transformation here should be (at least) restricted 
to respect the (tri-)periodicity of the orbit obtained 
from Eq.~\eqref{Heq-wJ}; see Sec. IV of Ref.~\cite{Fujita:2016igj} and 
Sec.~7.6 of Ref.~\cite{Barack:2018yvs} for more details.
}

In the context of the Hamiltonian formalism of Ref.~\cite{Fujita:2016igj}, 
the gauge freedom corresponds to the infinitesimal canonical transformation 
associated with a generating function $\Xi \equiv \xi^{\mu} u_{\mu}$ 
with a gauge vector $\xi^{\mu} = O(\eta)$, 
which describes a standard infinitesimal coordinate transformation, 
$x^{\mu} \to x^{\mu} + \xi^{\mu} + O(\eta^2)$.
Assuming that $\xi^{\mu} = O(\eta)$ holds everywhere in the spacetime 
(to avoid any spurious secular growth in the metric perturbation), 
the gauge transformation induces 
$J_{\alpha} \to J_{\alpha} + \hat \delta_\xi J_{\alpha}$, 
where $\hat \delta_\xi J_{\alpha}$ do not contain any secularly growing terms 
(see Eq.~(4.7) of Ref.~\cite{Fujita:2016igj}). 
By taking the $\tau$-average of the $\tau$-derivative of this relation, 
we have 
\begin{equation}\label{gauge-dotJ}
\hat \delta_\xi \langle {\dot J}_{\alpha} \rangle_\tau
= \lim_{T\to\infty}\frac{1}{2T} \left[
\hat\delta_\xi J_{\alpha}(T) - \hat\delta_\xi J_{\alpha}(-T)\right] 
=0\,, 
\end{equation}
and this result immediately establishes 
the quasi gauge invariance of $\langle {\dot J_{\alpha}} \rangle_\tau$.

\section{Flux formulae from the averaged Hamilton equation}
\label{sec:flux}

In this section, we specialize the discussion to the radiative sector 
of Eqs.~\eqref{Jdot-nonres} and~\eqref{Jdot-res}, and translate them 
into a more practical language of flux formulae. 
These formulae involve only asymptotic amplitudes of gravitational waves, 
which are readily computable in well-developed Teukolsky framework 
of black-hole perturbation theory 
(see, e.g.,~Refs.~\cite{Nakamura:1987zz,Sasaki:2003xr} for reviews). 
We shall leave for the future the more difficult task of 
the actual evaluation of the symmetric piece 
${\delta \langle \Hsym \rangle_\tau} / {\delta w^{\perp}_0}$
in Eq.~\eqref{res-dHsym} for the resonant orbits.~\footnote{
Preliminary work on the toy problem of the scalar-field self-force 
can be found in Ref.~\cite{Isoyama:2013yor}.}

Throughout this section we will use an overbar and ``c.c.'' 
to mean complex conjugate. 
Our strategy here closely follows a number of techniques developed 
by Sago et al.~\cite{Sago:2005gd,Sago:2005fn,Ganz:2007rf}, 
Drasco et al.~\cite{Drasco:2003ky,Drasco:2005is,Drasco:2005kz}, 
Grossman et al.~\cite{Grossman:2011im} 
and Flanagan et al.~\cite{Flanagan:2012kg}, 
and we adopt the conventions of Ref.~\cite{Flanagan:2012kg}.

\subsection{Radiative interaction Hamiltonian in terms of 
Teukolsky mode functions}
\label{subsec:Hrad-T}

A natural starting point of our computation may be 
the Green's-function-based expression for $\Hrad$ given by Eq.~\eqref{Hint-G}. 
The key notion here is that the tensorial radiative Green's function 
$G^{\mu\nu\,\rho\sigma}_{\mathrm {(rad)}}$ 
in particular traceless ``radiation'' gauges 
can be reconstructed from the scalar radiative Green's function 
for the Teukolsky equation of perturbed Weyl curvature scalars, 
as first derived by Gal'tsov~\cite{Galtsov:1982hwm} 
(and later corrected in Ref.~\cite{Drasco:2005is}).  
The scalar radiative Green's function can be expressed in terms of 
only its \textit{homogeneous solutions} to the Teukolsky equations---see 
Eq.~\eqref{Grad-Teukolsky}--- which are separated in all of the variables, 
and the reconstruction procedure is straightforward, 
relying on the classical method developed long ago 
by Chrzanowski~\cite{Chrzanowski:1975wv}  
and Cohen and Kegeles~\cite{Cohen:1974cm,Kegeles:1979an}.
As we have seen, since (the right-hand side of) Eqs.~\eqref{Jdot-nonres} 
and~\eqref{Jdot-res} are quasi gauge invariant, 
the reconstruction approach provides 
an extremely efficient route to evaluate them.~\footnote{
We should mention that the mass and angular momentum 
of the background Kerr spacetime are not altered 
(in the Abott-Deser sense~\cite{Abbott:1981ff}) 
by adding the metric perturbation associated 
with the reconstructed $G^{\mu\nu\,\rho\sigma}_{\mathrm {(rad)}}$ 
through the method of 
Ref.~\cite{Chrzanowski:1975wv,Cohen:1974cm,Stewart:1978tm,Kegeles:1979an}. 
References~\cite{Wald:1973JMAP,Merlin:2016boc,vandeMeent:2017fqk} discuss 
this statement in more detail.}

The derivation of $\Hrad$ based on the Teukolsky equations 
of Eqs.~\eqref{F-TeukolskyEqs}
is provided in Refs.~\cite{Galtsov:1982hwm}
as well as in Appendix~A of Ref.~\cite{Sago:2005fn}, 
and we shall not need the technical details here. 
Thus, we simply import the final result from Eq.~(3.9) of 
Ref.~\cite{Sago:2005fn}:~\footnote{
The retarded and advanced metric perturbations share 
the same static $\omega = 0$ modes, and they are not contained 
in $\Hrad$. Their contribution is classified here as part of $\Hsym$.} 
\begin{align}\label{F-Hrad}
\Hrad (x,\,u) &= - \int \rmd \omega \sum_{\ell m}
\frac{\mu}{4 i \omega^3}
\left\{
Z^{\mathrm {out}}_{\wlm} \Phi^{\mathrm {out}}_{\wlm} (x,\,u) 
+ 
\frac{\omega}{p_{\omega m}}
Z^{\mathrm {down}}_{\wlm} \Phi^{\mathrm {down}}_{\wlm} (x,\,u) 
\right\} + {\mathrm {(c. c. )}}\,,
\end{align}
where $\omega$ is a continuous frequency, 
$\sum_{\ell m} \equiv \sum_{\ell = 2}^{\infty} \sum_{m=-\ell}^{\ell}$ 
with a pair of integers $(\ell,\,m)$, 
$p_{\omega m} \equiv \omega - m a / (2 M r_+)$ 
(where $r_{+} \equiv M + \sqrt{M^2 - a^2}$)
is the superradiant factor, 
and the ``down'' and ``out'' modes are defined 
by specifying the boundary conditions for the Teukolsky functions 
imposed at the event horizon and infinity---see Eqs.~\eqref{mode2}. 
Here, we have defined a mode scalar $\Phi^{\mathrm {out/down}}_{\wlm}$
on the phase space (with the spin-weight $s = -2$) by~\footnote{
The choice of spin weights corresponds to that 
of one of two radiation gauges. 
For $s = -2$, the radiative metric perturbation (re)constructed 
from $G^{\mu\nu\,\rho\sigma}_{\mathrm {(rad)}}$ satisfies 
$h_{\mu \nu}^{\mathrm {(rad)}} n^{\mu} = 0 $
with a null tetrad $n^{\mu}$---see Eq.~\eqref{ln-v});  
the details are reviewed, e.g., in Ref.~\cite{Whiting:2005hr}.
}
\begin{equation}\label{def-Phi}
\Phi^{\mathrm {out/down}}_{\wlm} (x,\,u) 
\equiv 
{}_{-2} \tau^{\dagger}_{\mu \nu} \, 
{}_{-2} \Psi^{\mathrm {out/down}}_{\omega \ell m}(x) u^{\mu} u^{\nu} \,,
\end{equation}
in terms of a pure mode function (for the Heltz potential)
given---see Eq.~\eqref{F-PsiT}---by 
\begin{equation}\label{Heltz}
{}_{-2} \Psi^{\mathrm {out/down}}_{\omega \ell m} (x)
= 
\,_{-2} N_{\omega \ell m} \, {}_{2} R^{\mathrm {out/down}}_{\omega \ell m} (r) 
\, {}_{-2} S_{\omega \ell m} (\theta,\,\phi) e^{- i \omega t}\,,
\end{equation}
where 
${}_{-2} \tau^{\dagger}_{\mu \nu}$ is a certain second-order 
differential operator as in Eq.~\eqref{def-taud}, 
${}_{s} S_{\omega \ell m}$ is the spin-weighted spheroidal harmonics, 
${}_{-s} R^{\mathrm {out/down}}_{\omega \ell m}$ 
is a (spin-flipped) ``out'' or ``down'' mode 
of the radial Teukolsky function, 
and ${}_s N_{\omega \ell m}$ is a normalization factor. 
The amplitude of the mode scalar $\Phi^{\mathrm {out/down}}_{\wlm}$,
$Z^{\mathrm {out/down}}_{\wlm}$, 
is given by integrating the complex conjugate 
${\overline {\Phi^{\mathrm {out/down}}_{\wlm}}} (\tau)
= 
{\overline {\Phi^{\mathrm {out/down}}_{\wlm}}} (x(\tau),\,u(\tau))$  
along the orbit as
\begin{equation}\label{def-Z}
Z^{\mathrm {out/down}}_{\wlm}
\equiv
\int \rmd \tau' \, {\overline {\Phi^{\mathrm {out/down}}_{\wlm}}} (\tau')\,.  
\end{equation} 
The explicit expressions for ${}_{-2} \tau^{\dagger}_{\mu \nu}$ 
and 
$\Phi^{\mathrm {out/down}}_{\wlm}$ are provided 
in Eq.~\eqref{def-taud} in Appendix~\ref{sec:Teukolsky} 
and Eq.~(A.55) of Ref.~\cite{Sago:2005fn}, respectively.

\subsection{Harmonic decomposition of the mode scalars and amplitudes}
\label{subsec:harmonic}

When inserting Eq.~\eqref{F-Hrad} into the $\tau$-averaged Hamilton equation 
of Eqs.~\eqref{Jdot-nonres} and~\eqref{Jdot-res}, we find 
\begin{align}\label{F-dJdt}
\langle \dot J_{\alpha} \rangle_\tau
&= \int \rmd \omega \sum_{\ell m}
\frac{\mu}{4 i \omega^3}
\left\{
Z^{\mathrm {out}}_{\wlm} 
\left \langle \left(
\frac{\partial \Phi^{\mathrm {out}}_{\wlm}} {\partial w^{\alpha}} 
\right)_{J} \right \rangle_\tau
+ 
\frac{\omega}{p_{\omega m}}Z^{\mathrm {down}}_{\wlm} 
\left \langle \left(
\frac{\partial \Phi^{\mathrm {down}}_{\wlm}} {\partial w^{\alpha}} 
\right)_{J} \right \rangle_\tau 
\right\} \cr 
\quad &+ {\mathrm {(c. c. )}}\,. 
\end{align}
In this subsection, we shall simplify this expression, 
first for the non-resonant case and
next for the resonant case, respectively. 
We will consider only the ``out'' mode of Eq.~\eqref{F-dJdt} below; 
the ``down'' mode is precisely analogous. 

The mode scalar $\Phi^{\mathrm {out}}_{\wlm}$ defined 
on the stationary and axially symmetric Kerr background is proportional 
to $e^{- i \omega t} e^{i m \phi}$, which implies 
$\Phi^{\mathrm {out}}_{\wlm} \propto e^{- i \omega w^t} e^{i m w^{\phi}}$ 
from Eq.~\eqref{def-w}. 
Using this result,
$\Phi^{\mathrm {out}}_{\wlm}$
allows the Fourier expansions in $w^{r}$ and $w^{\theta}$ given by 
\begin{equation}\label{F-Phi}
\Phi^{\mathrm {out}}_{\wlm}(w,\,J)
= 
\sum_{k,n} 
\phi^{\mathrm {out}}_{\wlmkn}(J) 
e^{-i(\omega w^t - m w^{\phi} - k w^\theta - n w^r)}\,,
\end{equation}
where the sum 
$\sum_{k,n} = \sum_{k = -\infty}^{+\infty} \sum_{n = -\infty}^{+\infty}$
is over pairs of integers $(k,\, n)$, 
and the Fourier coefficients are 
\begin{equation}\label{F-phi}
{\phi}^{\mathrm {out}}_{\omega \ell m k n}(J) 
=
\frac{1}{(2 \pi)^2}
\int_0^{2 \pi} \rmd w^r \int_0^{2 \pi} \rmd w^\theta 
e^{-i(k w^{\theta} + n w^r)} \Phi^{\mathrm {out}}_{\wlm}(w,\,J)\,.
\end{equation}
It is important to recognize that 
$\Phi^{\mathrm {out}}_{\wlm}$ is a function defined
{on the phase space},  
not restricted to the orbit accounting for Eq.~\eqref{norm-wJ}. 
Inserting Eqs.~\eqref{soln-w} and~\eqref{F-Phi} into Eq.~\eqref{def-Z}, 
a straightforward computation returns 
\begin{align}\label{F-Z}
Z^{\mathrm {out}}_{\wlm}
&=
\sum_{k,n}
\left.
\int \rmd \tau\, 
{\overline {\phi^{\mathrm {out}}_{\omega \ell m k n}}}(J)
e^{i(\omega w^t - m w^{\phi} - k w^{\theta} - n w^r)} 
\right|_\gamma \cr
&= 
\sum_{k,n} {\tilde Z}_{\ell m k n}^{\mathrm {out}} \,e^{i \chi_{m k n}} 
\delta (\omega - \omega_{m k n}) \,.  
\end{align}
Here, we have introduced discretized fundamental frequencies, 
initial phases, and amplitudes as
\begin{subequations}\label{def-wc}
\begin{align}
\label{w-mkn}
\omega_{m k n} 
&\equiv 
m \Omega^{\phi} + k \Omega^{\theta} + n \Omega^{r}\,,
\\
\label{c-mkn}
\chi_{m k n}
&\equiv
\omega_{m k n} w^t_{0} 
- (m w^\phi_{0} + k w^{\theta}_{0} + n w^{r}_{0})\,,
\\
\label{Z-lmkn}
{\tilde Z}_{\ell m k n}^{\mathrm {out}} 
&\equiv \left. 2 \pi z\, 
{\overline {\phi^{\mathrm {out}}_{\omega_{m k n} \ell m k n}}}(J)
\right|_\gamma \,,
\end{align}
\end{subequations}
respectively, replacing $\omega$ with $\omega_{mkn}$ 
because of the delta function in Eq.~\eqref{F-Z}.
Notice that the amplitude $Z^{\mathrm {out}}_{\wlm}$ depends 
on the initial phases of the orbit only through
the overall phase $e^{i \chi_{m k n}}$~\cite{Drasco:2003ky,Drasco:2005kz}. 

We may now simplify the $\tau$-averaged derivative 
of $\Phi^{\mathrm {out}}_{\wlm}$ with respect to $w^{\alpha}$ 
in Eq.~\eqref{F-dJdt}. 
Invoking Eqs.~\eqref{soln-w},~\eqref{F-Phi}, and~\eqref{F-Z}, we have 
\begin{align}\label{dPhidw}
& \left. 
\int \rmd \omega\, Z^{\mathrm {out}}_{\wlm}
\left( 
\frac{\partial \Phi^{\mathrm {out}}_{\wlm}}{\partial w^{\alpha}}
\right)
\right|_{\gamma} \cr 
& \quad = 
\sum_{k,n} \sum_{k',n'} \!
{\tilde Z}_{\ell m k n}^{\mathrm {out}} \,e^{i \chi_{m k n}} 
\,(i {\varepsilon}_{\alpha}) \!
\left. 
\phi^{\mathrm {out}}_{\omega_{m k n} \ell m k' n'}(J)
\,e^{-i(\omega_{m k n} w^t - m w^{\phi} - k' w^\theta - n' w^r)} 
\right|_{\gamma}
\cr 
& \quad = 
i
\sum_{k,n} \sum_{k', n'}  \!
\left. 
{\tilde Z}_{\ell m k n}^{\mathrm {out}} 
\left(
{\varepsilon}_{\alpha}
\phi^{\mathrm {out}}_{\omega_{m k n} \ell m k' n'}(J) \,\right) 
e^{i \{ (k' - k) w^{\theta} + (n' - n) w^r \}}
\right|_{\gamma}
\,, 
\end{align}
where 
\begin{equation}\label{vareps}
{\varepsilon}_{\alpha} 
\equiv 
\left(-\omega_{m k n},\,n',\,k',\,m \right)\, 
\end{equation}
and we have used Eqs.~\eqref{def-wc}. 
The expression of Eq.~\eqref{dPhidw} applies to 
both non-resonant and resonant orbits. 
In the non-resonant case, the $\tau$-average of Eq.~\eqref{dPhidw} forces 
$k' = k$ and $n' = n$ in the sum, and we just arrive at 
\begin{equation}\label{dPhidw-ave}
\int \rmd \omega\, Z^{\mathrm {out}}_{\wlm} 
\left \langle
\left( 
\frac{\partial \Phi^{\mathrm {out}}_{\omega \ell m}}{\partial w^{\alpha}}
\right)_{\,J}
\right \rangle_\tau
= 
\frac{i}{2 \pi z} \sum_{k,n}
{\varepsilon}_{\alpha}\, | {\tilde Z}_{\ell m k n}^{\mathrm {out}} |^2\,,
\end{equation} 
and the initial phase $w_0^{\alpha}$ of the orbit does not appear here. 

In the resonant case, however, all pairs $(k,\,n)$ satisfy 
$k \Omega^{\theta} + n \Omega^r = N \tilde{\Omega}$ with
an integer $N \equiv k \beta^{\theta} + n \beta^{r}$ 
owing to the resonant condition of Eq.~\eqref{res-Omega}. 
In terms of the angle variables $w^r$ and $w^{\theta}$, this implies 
\begin{align}\label{w-res}
k w^{\theta}(\tau) + n w^r (\tau) 
&= 
N \tilde{\omega} \,(\tau - \tau_0) 
+ 
k w^{\theta}_0 + n w^{r}_0\, \cr 
&= 
N \tilde{\omega} \,(\tau - \tau_0) 
+ 
\frac{N}{\beta^r}w^{r}_0 + k \beta^{\theta} w^{\perp}_0 \,, 
\end{align}
with $\tilde{\omega} \equiv \tilde{\Omega} / z$ and Eq.\eqref{D-perp}.
Then, the $\tau$-average of Eq.~\eqref{dPhidw}
enforces $N' = N$ (where $N' \equiv k' \beta^{\theta} + n' \beta^{r}$)
in the sum, 
and a straightforward computation, making use of the second equality 
of Eq.\eqref{w-res}, returns 
\begin{align}\label{dPhidw-ave-res}
& \int \rmd \omega\, Z^{\mathrm {out}}_{\wlm} 
\left \langle
\left( 
\frac{\partial \Phi^{\mathrm {out}}_{\omega \ell m}}{\partial w^{\alpha}}
\right)_{\,J}
\right \rangle_\tau \cr
& \quad = 
\frac{i}{2 \pi z} \sum_{N=-\infty}^{\infty} 
\Biggl(
\sum_{({k}, {n})_{N}}  
{\tilde Z}^{\mathrm {out}}_{\ell m (k n)} 
e^{- i k \beta^{\theta} w^{\perp}_0} 
\Biggr)
\Biggl(
\sum_{({k'}, {n'})_{N}} {\varepsilon}_{\alpha}\, 
{\overline {{\tilde Z}^{\mathrm {out}}_{\ell m (k' n') }}} 
e^{ i k' \beta^{\theta} w^{\perp}_0} 
\Biggr)\,,
\end{align}
where ${\varepsilon}_{\alpha} = (-\omega_{m N},\,n',\,k',\,m)$ now, and 
\begin{subequations}\label{def-wc-res}
\begin{align}\label{Z-lmkn-res}
{\tilde Z}_{\ell m (k n)}^{\mathrm {out}} 
&\equiv \left. 2 \pi z\, 
{\overline {\phi^{\mathrm {out}}_{\omega_{m N} \ell m k n}}}(J)
\right|_\gamma \,,
\\
\label{w-mN}
\omega_{m N} 
&\equiv 
m \Omega^{\phi} + N \Omega \,, 
\end{align}
\end{subequations}
respectively. 
The notation $(k, n)_N$ means the summation over all pairs 
of $(k, n)$ which satisfy the relation $N = k \beta^{\theta} + n \beta^{r}$. 
Notice that the cross term between pairs $(k, n)_N$ and $(k', n')_N$ 
remains even if $k \neq k'$ and $n \neq n'$.
It is important to recognize that, unlike Eq.~\eqref{dPhidw-ave},
the expression of Eq.~\eqref{dPhidw-ave-res} can generically depend 
on the initial phase through the resonant phase $w^{\perp}_0$.
Indeed, the resonant phase term $(k - k') \beta^{\theta} w^{\perp}_0$ 
of Eq.~\eqref{dPhidw-ave-res} vanishes only 
after the average over $w^{\perp}_0$.  
These key observations agree with the results found 
in Refs.~\cite{Grossman:2011im,Flanagan:2012kg}.

\subsection{Flux formulae}
\label{subsec:fluxes}

Our final task is to assemble the results in previous subsections, 
and derive flux formulae from the $\tau$-averaged Hamilton equations 
of Eq.~\eqref{F-dJdt}.
For the non-resonant case, Eq.~\eqref{dPhidw-ave} can now be substituted 
into the right-hand side of Eq.~\eqref{F-dJdt}. 
We simply arrive at 
\begin{equation}\label{Flux-nonres}
\left \langle 
\frac{\rmd J_{\alpha}}{\rmd t} 
\right \rangle_t 
=
- \mu \sum_{\ell m k n} 
\frac{{\varepsilon}_{\alpha}}{4 \pi \omega^3_{m k n}}
\left(
| {\tilde Z}_{\ell m k n}^{\mathrm {out}} |^2
+
\frac{\omega_{m k n}}{p_{m k n}} 
| {\tilde Z}_{\ell m k n}^{\mathrm {down}} |^2 
\right)\,, 
\end{equation}
where we have used the relation about the various long-time averages 
(see, e.g., Sec.~9 of Ref.~\cite{Drasco:2005is}) 
\begin{equation}\label{convert-aveJ}
\left \langle \frac{\rmd J_\alpha}{\rmd t} \right \rangle_t 
=
z \left \langle \frac{\rmd J_\alpha}{\rmd \tau} \right \rangle_\tau\,,
\end{equation}
with the redshift variable $z$---recall Eq.~\eqref{def-omega}. 
Equation~\eqref{Flux-nonres}
is the final form of the flux formulae for non-resonant orbits.
For the resonant case, recalling Eq.~\eqref{dPhidw-ave-res}, 
it is convenient to introduce 
the initial-phase-dependent amplitudes of ``out'' mode defined 
by~\cite{Grossman:2011im,Flanagan:2012kg} 
\begin{align}\label{cal-XYZ}
\left\{ 
\begin{array}{l} 
{\tilde {\cal X}}_{\ell m N}^{\mathrm {out}} \\
{\tilde {\cal Y}}_{\ell m N}^{\mathrm {out}} \\
{\tilde {\cal Z}}_{\ell m N}^{\mathrm {out}} \\
\end{array} 
\right\}
\equiv
\sum_{(k,n)_N} 
\left\{ 
\begin{array}{l} 
n \\ 
k \\
1 \\
\end{array} 
\right\}
\, e^{- i k \beta^{\theta} w^{\perp}_0} {\tilde Z}_{\ell m k n}^{\mathrm {out}}\,, 
\end{align}
and those of the ``down'' modes as well.
Then, Eq.~\eqref{dPhidw-ave-res} can be 
substituted into the right-hand side of Eq.~\eqref{F-dJdt}. 
By using Eqs.~\eqref{Jdot-res} and~\eqref{res-dHsym},
a simple computation with the relation of Eq.~\eqref{convert-aveJ}, 
now recovering the symmetric contribution 
to Eq.~\eqref{F-dJdt} for completeness, gives 
\begin{subequations}\label{Flux-res}
\begin{align}
\label{dJtdt-res}
\left \langle 
\frac{\rmd J_{t}}{\rmd t} 
\right \rangle_t 
&=
\mu \sum_{\ell m N} 
\frac{1}{4 \pi \omega^2_{m N}}
\left(
| {\tilde {\cal Z}}_{\ell m N}^{\mathrm {out}} |^2
+
\frac{\omega_{m N}}{p_{m N}} 
| {\tilde {\cal Z}}_{\ell m N}^{\mathrm {down}} |^2 
\right)\,,
\\
\label{dJphidt-res}
\left \langle 
\frac{\rmd J_{\phi}}{\rmd t} 
\right \rangle_t 
&=
-\mu \sum_{\ell m N} 
\frac{m}{4 \pi \omega^3_{m N}}
\left(
| {\tilde {\cal Z}}_{\ell m N}^{\mathrm {out}} |^2
+
\frac{\omega_{m N}}{p_{m N}} 
| {\tilde {\cal Z}}_{\ell m N}^{\mathrm {down}} |^2 
\right)\,,
\\
\label{dJrdt-res}
\left \langle 
\frac{\rmd J_{r}}{\rmd t} 
\right \rangle_t 
&= 
-\mu \sum_{\ell m N} 
\frac{1}{4 \pi \omega^3_{m N}}
\left(
{\mathrm {Re}} [
{\overline {{\tilde {\cal X}}_{\ell m N}^{\mathrm {out}}}} 
{\tilde {\cal Z}}_{\ell m N}^{\mathrm {out}} ] 
+
\frac{\omega_{m N}}{p_{m N}} 
{\mathrm {Re}} [
{\overline {{\tilde {\cal X}}_{\ell m N}^{\mathrm {down}}}} 
{\tilde {\cal Z}}_{\ell m N}^{\mathrm {down}} ] 
\right) 
+
\frac{z}{2 \beta^{r}} 
\frac{\delta \left\langle \Hsym \right\rangle_\tau} {\delta w^{\perp}_0}\,,
\\
\label{dJthdt-res}
\left \langle 
\frac{\rmd J_{\theta}}{\rmd t} 
\right \rangle_t 
&= -\mu \sum_{\ell m N} 
\frac{1}{4 \pi \omega^3_{m N}}
\left(
{\mathrm {Re}} [
{\overline {{\tilde {\cal Y}}_{\ell m N}^{\mathrm {out}}}}   
{\tilde {\cal Z}}_{\ell m N}^{\mathrm {out}} ] 
+
\frac{\omega_{m N}}{p_{m N}} 
{\mathrm {Re}} [
{\overline {{\tilde {\cal Y}}_{\ell m N}^{\mathrm {down}}}}   
{\tilde {\cal Z}}_{\ell m N}^{\mathrm {down}} ] 
\right)
-
\frac{z}{2 \beta^{\theta}} 
\frac{\delta \left\langle \Hsym \right\rangle_\tau} 
{\delta w^{\perp}_0}\,,
\end{align}
\end{subequations} 
which are the final forms of the flux formulae for resonant orbits. 

In general, the flux formulae of 
$\langle \rmd J_{r} / \rmd t \rangle_t$ and 
$\langle \rmd J_{\theta} / \rmd t \rangle_t$ in Eqs.~\eqref{Flux-res} 
are invalid unless one includes
${\delta \langle \Hsym \rangle_\tau}/{\delta w^{\perp}_0}$.
Nevertheless, we see that the special linear combination 
\begin{align}\label{dJbdt-res}
\left \langle 
\frac{\rmd J_{\parallel}}{\rmd t} 
\right \rangle_t 
&\equiv
\beta_r
\left \langle 
\frac{\rmd J_{r}}{\rmd t} 
\right \rangle_t 
+
\beta_\theta
\left \langle 
\frac{\rmd J_{\theta}}{\rmd t} 
\right \rangle_t \cr 
&=
- \mu \sum_{\ell m N} 
\frac{N}{4 \pi \omega^3_{m N}}
\left(
| {\tilde {\cal Z}}_{\ell m N}^{\mathrm {out}} |^2
+
\frac{\omega_{m N}}{p_{m N}} 
| {\tilde {\cal Z}}_{\ell m N}^{\mathrm {down}} |^2 
\right)\,,
\end{align}
\textit{is} valid because it does not involve any contribution 
from ${\delta \langle \Hsym \rangle_\tau}/{\delta w^{\perp}_0}$ 
at all, thanks to Eqs.~\eqref{res-Hsym} and 
$N = k \beta^{\theta} + n \beta^{r}$. 
This expression is also useful since it is 
expressed in terms of 
${\tilde {\cal Z}}_{\ell m N}^{\mathrm {out, down}}$ only, 
in the same manner as Eqs.~\eqref{dJtdt-res} and~\eqref{dJphidt-res}. 
We expect that a most advanced self-force code 
(such as the one of Ref.~\cite{vandeMeent:2018rms})
will soon be able to test this relation directly.

\section{Flux formulae for the energy, angular momentum, and Carter constant}  
\label{sec:dPdt}

In this section we calculate the canonical transformation 
between the evolution of the specific energy, 
azimuthal angular momentum and Carter constant ${\dot P}_{\alpha}$--- recall 
Eq.~\eqref{def-P}---and that of the action variables ${\dot J}_{\alpha}$ 
using Eq.~\eqref{JvsP}, and produce
the flux formulae for $\langle {\rmd P}_{\alpha} / {\rmd t} \rangle_t$ 
in terms of $\langle {\rmd J}_{\alpha} / {\rmd t} \rangle_t$ 
obtained in Sec.~\ref{sec:flux}. 
To prepare the way for the discussion, 
we introduce the (Carter-)Mino time~\cite{Carter:1968rr,Mino:2003yg} $\lambda$ 
related to the proper time $\tau$ of Eq.~\eqref{norm-xu} by 
\begin{equation}\label{Mino-t}
\rmd \lambda \equiv \Sigma^{-1} \rmd \tau + O(\eta) \,,
\end{equation}
and the associated long-time averaging 
$\langle f(\lambda)\rangle_{\lambda}$ for various functions $f(\lambda)$; 
recall Eq.~\eqref{longtime}. 

\subsection{Identities for Kerr geodesics}
\label{subsec:dJdP}

We establish here a number of identities satisfied 
by the partial derivatives of the action variables $J_{\alpha}$ 
with respect to the canonical momenta $P_{\alpha}$ 
in the test-mass limit $\eta \to 0$. 
The notation
for the expression of Kerr geodesics in terms of $\lambda$ 
is adopted from Drasco and Hughes~\cite{Drasco:2003ky} 
and Fujita and Hikida~\cite{Fujita:2009bp}. 

We begin by computing the partial derivatives 
$(\partial J_r / \partial {\hat Q})$ and 
$(\partial J_\theta / \partial {\hat C})$. 
From the definitions of actions of Eqs.~\eqref{def-J0}, 
we immediately---see, e.g., Eqs.(3), (5), and (7) 
of Ref.~\cite{Fujita:2009bp}---obtain
\begin{subequations}\label{dJdQ}
 \begin{align}
 \left( \frac{\partial J_r}{\partial {\hat Q}} \right)
 &= 
 -\frac{1}{2 \pi} \int^{r_{\mathrm {max}}}_{r_{\mathrm{\min}}} 
 \frac{\rmd r}{\sqrt{R}} 
 = 
 - \frac{1}{2 \Upsilon^r}\,,
\\
\label{dJdC}
 \left( \frac{\partial J_\theta}{\partial {\hat C}} \right) 
 &= 
 \frac{1}{\pi} \int^{\cos \theta_{\mathrm {min}}}_{0} 
 \frac{\rmd \cos \theta}{\sin \theta\, \sqrt{\Theta}} 
 = 
 \frac{1}{2 \Upsilon^\theta}\,, 
 \end{align}
\end{subequations}
where $r_{\mathrm {min/max}} (r_{\mathrm {min}} \leq r_{\mathrm {max}})$ are 
the two largest roots of $R(r,\,P) = 0$, $\cos \theta_{\mathrm {min}} >0$ 
is the smallest positive root of $\Theta (\cos \theta,\, P) = 0$, recall 
Eqs.~\eqref{def-R-Theta}, and 
\begin{equation}\label{def-U}
\Upsilon^{\alpha} 
\equiv
\left( \Gamma,\,\Upsilon^r,\,\Upsilon^\theta,\,\Upsilon^\phi \right)\,,
\end{equation}
are the angular frequencies of Kerr geodesics 
defined with respect to the Mino time $\lambda$.
Note that we have 
$({\partial r_{\mathrm {min}}}/\partial P_{\alpha}) \sqrt{R} 
= 0 = 
({\partial r_{\mathrm {max}}}/\partial P_{\alpha}) \sqrt{R}$ 
because $R(r,\,P) =0$ at $r = r_{\mathrm {min/max}}$, 
(and similarly for 
$(\partial \cos \theta_{\mathrm {min}}/\partial P_{\alpha}) \sqrt{\Theta}= 0$).
Equations~\eqref{dJdQ} are clearly separated in $r$ and $\theta$.
This comes from the separation property of the generating function 
$W(x,\,P)$ in Eq.~\eqref{def-W}, 
and it reflects the fact that the Carter constant 
${\hat Q}$ (or ${\hat C}$) is the separation constant 
for the Hamilton-Jacobi equation of $H^{(0)}$
in Eq.~\eqref{H0H1}~\cite{Carter:1968rr,Schmidt:2002qk}. 

Next, we evaluate $(\partial J_{r/\theta} / \partial {\hat E})$ 
and $(\partial J_{r/\theta} / \partial {\hat L})$, respectively. 
A similar calculation (recall that 
${\hat P} = {\hat E} (r^2 + a^2) -  a {\hat L}$, and  
see, e.g., Eqs. (1) and (16) of Ref.~\cite{Fujita:2009bp}) reveals that 
\begin{subequations}\label{dJdEL}
 \begin{align}
 \label{dJrdE}
 \Upsilon_r \, \left( \frac{\partial J_r}{\partial {\hat E}} \right)_{\! {\hat Q}} 
 &= 
 \frac{2}{\Lambda_r} \int^{r_{\mathrm {max}}}_{r_{\mathrm{\min}}} 
 \frac{\rmd r}{\sqrt{R}} 
 \left( 
 \frac{r^2 + a^2}{\Delta} {\hat P}
 \right)\,,
\\ 
\label{dJrdL}
 \Upsilon_r \, \left( \frac{\partial J_r}{\partial {\hat L}} \right)_{\! {\hat Q}} 
 &= - 
 \frac{2}{\Lambda_r} \int^{r_{\mathrm {max}}}_{r_{\mathrm{\min}}} 
 \frac{\rmd r}{\sqrt{R}} 
 \left( 
 \frac{a}{\Delta} {\hat P}
 \right)\,, 
\\ 
\label{dJthdE}
 \Upsilon_\theta \, \left( \frac{\partial J_\theta}{\partial {\hat E}} \right)_{\! {\hat C}} 
 &= 
 \frac{4}{\Lambda_\theta} \int^{\cos \theta_{\mathrm {min}}}_{0} 
 \frac{\rmd \cos \theta}{\sin \theta\,\sqrt{\Theta}} \, 
 \left( a^2 {\hat E} \cos^2 \theta \right)\,,
\\
\label{dJthdL}
 \Upsilon_\theta \, \left( \frac{\partial J_\theta}{\partial {\hat L}} \right)_{\! {\hat C}} 
 &= 
 -\frac{4}{\Lambda_\theta} \int^{\cos \theta_{\mathrm {min}}}_{0} 
 \frac{\rmd \cos \theta}{\sin \theta\, \sqrt{\Theta}} 
 \left( {\hat L} \cot^2 \theta \right)\,, 
 \end{align}
\end{subequations}
where we have used the relation $\Upsilon_r = 2 \pi / \Lambda_r$ 
and $\Upsilon_\theta = 2 \pi / \Lambda_{\theta}$ 
with the Mino-time radial period $\Lambda_r$ 
and polar period $\Lambda_{\theta}$. 
Recalling the geodesic equation in terms of $\lambda$
(see, e.g., Eqs. (2) of Ref.~\cite{Fujita:2009bp}), we have 
\begin{equation}\label{non-res}
\rmd \lambda 
= 
\frac{\rmd r}{\sqrt{R}} 
=
\frac{\rmd \cos \theta}{\sin \theta \sqrt{\Theta}}\,.  
\end{equation}
This relation implies that the right-hand sides of Eqs.~\eqref{dJdEL} 
are averaged over orbital periods $\Lambda_r$ or $\Lambda_{\theta}$ 
(i.e., $\Lambda_r^{-1} \int_0^{\Lambda_r} \rmd \lambda \cdots$ 
or $\Lambda_{\theta}^{-1} \int_0^{\Lambda_{\theta}} \rmd \lambda \cdots$), 
and they can be replaced by the equivalent long-time average over $\lambda$ 
of Eq.~\eqref{longtime}~\footnote{
This replacement is valid for both the non-resonant 
and resonant cases because the right-hand sides of Eqs.~\eqref{dJdEL} 
are only functions of either $r$ or $\theta$.
}. 
We therefore end up with identities 
\begin{equation}\label{dJrdEL-ave}
 \left( \frac{\partial J_r}{\partial {\hat Q}} \right)^{-1}
 \left( \frac{\partial J_r}{\partial {\hat E}} \right)_{\! {\hat Q}}  
 = -
 2 \left \langle  
 \frac{r^2 + a^2}{\Delta} {\hat P}
 \right \rangle_{\lambda}\,,  
 \quad 
 \left( \frac{\partial J_r}{\partial {\hat Q}} \right)^{-1} 
 \left( \frac{\partial J_r}{\partial {\hat L}} \right)_{\! {\hat Q}}  
 = 
 2 \left \langle \frac{a}{\Delta} {\hat P} \right \rangle_\lambda \,
\end{equation}
and 
\begin{equation}\label{dJthdEL-ave}
 \left( \frac{\partial J_\theta}{\partial {\hat C}} \right)^{-1} 
 \left( \frac{\partial J_\theta}{\partial {\hat E}} \right)_{\! {\hat C}}  
 = 
2 \left \langle a^2 {\hat E} \cos^2 \theta \right \rangle_\lambda \,, 
 \quad
 \left( \frac{\partial J_\theta}{\partial {\hat C}} \right)^{-1} 
 \left( \frac{\partial J_\theta}{\partial {\hat E}} \right)_{\! {\hat C}} 
 = 
 -2 \left \langle {\hat L} \cot^2 \theta \right \rangle_\lambda  \,,
\end{equation}
where we have used Eqs.~\eqref{dJdQ}.

\subsection{Averaged evolution of the energy, angular momentum 
and Carter constant} 
\label{subsec:C(J)}

First, we derive the expressions for ${\dot P}_{\alpha}$ 
in terms of ${\dot J}_{\alpha}$ using the canonical transformation 
of Eq.~\eqref{JvsP}. 
The evolution of the specific energy and angular momentum,
${\dot {\hat E}} = - {\dot J_t}$
and ${\dot {\hat L}} = {\dot J_\phi}$
trivially follow from their definitions in Eqs.~\eqref{def-J0}. 
The expressions for the evolution of the specific Carter 
constants, ${\dot {\hat {Q}}}$ and ${\dot {\hat {C}}}$ are 
easily produced from the proper-time derivative of 
$J_{r} = J_{r} (H^{(0)},\,{\hat E},\,{\hat L},\,{\hat {Q}})$
and 
$J_\theta = J_\theta (H^{(0)},\,{\hat E},\,{\hat L},\,{\hat {C}})$. 
After some simple algebra, making use of the identities 
Eqs.~\eqref{dJdQ},~\eqref{dJrdEL-ave} and~\eqref{dJthdEL-ave}, 
we arrive at 
\begin{subequations}\label{dotQC}
 \begin{align}\label{dotQ}
 \frac{1}{2} {\dot {\hat Q}}
 &= 
\left \langle \frac{r^2 + a^2}{\Delta} {\hat P} \right \rangle_\lambda
{\dot {\hat E}} 
-
\left \langle \frac{a}{\Delta} {\hat P} \right \rangle_\lambda
{\dot {\hat L}}
-
\Upsilon_{r} 
\left\{
{\dot {J_r}} 
- 
\left( \frac{\partial J_r}{\partial H^{(0)}} \right) {\dot {H}}^{(0)}
\right\}
\,,
\\
\label{dotC}
\frac{1}{2} {\dot {\hat C}} 
&=- 
\left \langle a^2 {\hat E} \cos^2 \theta \right \rangle_\lambda
{\dot {\hat E}} 
+
\left \langle L \cot^2 \theta \right \rangle_\lambda 
{\dot {\hat L}} 
+
\Upsilon_{\theta} 
\left\{
{\dot {J_\theta}} 
- 
\left( \frac{\partial J_\theta}{\partial H^{(0)}} \right) {\dot {H}}^{(0)}
\right\} \,. 
 \end{align}
\end{subequations}

We will now simplify ${\dot P}_{\alpha}$ averaged over $\tau$ 
to derive the associated flux formulae. 
Again, the averaged rate of change of ${\hat E}$ and ${\hat L}$ 
are trivially given by 
\begin{equation}\label{dELdt-ave}
\left \langle 
\frac{\rmd {\hat E}}{\rmd t} 
\right \rangle_t 
= - 
\left \langle 
\frac{\rmd J_{t}}{\rmd t} 
\right \rangle_t\,, 
\quad
\left \langle 
\frac{\rmd {\hat L}}{\rmd t} 
\right \rangle_t 
= 
\left \langle 
\frac{\rmd J_{\phi}}{\rmd t} 
\right \rangle_t\,, 
\end{equation}
making use of the relation of Eq.~\eqref{convert-aveJ}.
To simplify Eqs.~\eqref{dotQC}, 
recall that Eqs.~\eqref{defH} and~\eqref{norm-xu} imply that 
${\dot H} = {\dot {H}}^{(0)} + {\dot H}^{(1)} = 0$, which means, 
recalling the computation of Eq.~\eqref{Hsym-ave}, 
\begin{equation}\label{tauave-H1}
\left \langle {\dot {H}}^{(0)}  \right \rangle_\tau
= 
- \left \langle {\dot H}^{(1)} \right \rangle_\tau 
=
- \lim_{T\rightarrow\infty} 
\frac{1}{2T} \left[ {\Hint (T) - \Hint (-T)} \right] = 0\,, 
\end{equation}
for the $\tau$-averaging. 
With this relation, the $\tau$-averages of Eqs.~\eqref{dotQC} 
then read
\begin{subequations}\label{dotQC-ave}
 \begin{align}\label{dotQ-ave}
\frac{1}{2}
\left \langle \frac{\rmd {\hat Q}}{\rmd t} \right \rangle_t 
&=- 
\left \langle \frac{r^2 + a^2}{\Delta} {\hat P} \right \rangle_\lambda
\left \langle \frac{\rmd J_t}{\rmd t} \right \rangle_t 
-
\left \langle \frac{a}{\Delta} {\hat P} \right \rangle_\lambda
\left \langle \frac{\rmd J_\phi}{\rmd t} \right \rangle_t
-
\Upsilon_{r}
\left \langle \frac{\rmd J_{r}}{\rmd t} \right \rangle_t\,,
\\ 
\label{dotC-ave}
\frac{1}{2}
\left \langle \frac{\rmd {\hat C}}{\rmd t} \right \rangle_t 
&=  
\left \langle a^2 {\hat E} \cos^2 \theta \right \rangle_\lambda
\left \langle \frac{\rmd J_t}{\rmd t} \right \rangle_t 
+
\left \langle {\hat L} \cot^2 \theta \right \rangle_\lambda
\left \langle \frac{\rmd J_\phi}{\rmd t} \right \rangle_t
+ 
\Upsilon_{\theta}
\left \langle \frac{\rmd J_{\theta}}{\rmd t} \right \rangle_t\,,  
 \end{align}
\end{subequations}
where we have used Eqs.~\eqref{convert-aveJ} and~\eqref{dELdt-ave}. 

We can clearly see the $r$-$\theta$ split in the averaged rate of change 
of the Carter constants: the first version of the expression 
$\langle {\rmd {\hat Q}}/{\rmd t} \rangle_t$ in Eq.~\eqref{dotQ-ave} 
is described by only the ``$r$-components'' 
(i.e., $r,\,\Upsilon_r,\,\langle {\rmd J_r}/{\rmd t} \rangle_t$, etc.), 
and the second version of the expression 
$\langle {\rmd {\hat C}}/{\rmd t} \rangle_t$ in Eq.~\eqref{dotC-ave}
involves only the ``$\theta$-components'' 
(i.e., $\theta,\,\Upsilon_\theta,\,
\langle {\rmd J_\theta}/{\rmd t} \rangle_t$, etc.). 
However, they are not independent formulae because of the relation 
of Eq.~\eqref{C}, and indeed, Eq.~\eqref{dotQ-ave} is equivalent 
to Eq.~\eqref{dotC-ave}. 
This statement is easily understood, making use of the relation 
obtained from ``the first law of binary
mechanics''~\cite{Friedman:2001pf,LeTiec:2011ab,Fujita:2016igj}.
Importing, for example, from Sec.~5 of Ref.~\cite{Tiec:2013kua}, 
and keeping in mind that the fundamental frequency $\Omega^{\alpha}$ 
of Eq.~\eqref{def-omega} is related to the Mino-time frequency 
$\Upsilon^{\alpha}$ by the relation 
$\Omega^{\alpha} = \Upsilon^{\alpha} / \Gamma$, we have~\footnote{
Interestingly, the first-law relation of Eq.~\eqref{1stlaw-Sago} 
was implicitly derived in Refs.~\cite{Sago:2005gd,Sago:2005fn} 
(assuming the flux formulae of Eqs.~\eqref{Flux-nonres}); 
see, e.g., an unnumbered identity displayed below Eq. (3.25) 
of Ref.~\cite{Sago:2005fn}.
}
\begin{equation}\label{1stlaw-Sago}
\Gamma 
\left \langle \frac{\rmd {\hat {E}}}{\rmd t} \right \rangle_t 
=
\Upsilon^\phi\,
\left \langle \frac{\rmd {\hat L}}{\rmd t} \right \rangle_t 
+
\Upsilon^\theta\, \left \langle \frac{\rmd J_\theta}{\rmd t} \right \rangle_t 
+
\Upsilon^r\, \left \langle \frac{\rmd J_r}{\rmd t} \right \rangle_t\,, 
\end{equation}
with (see, e.g., Eqs.(7) of Ref.~\cite{Fujita:2009bp})
\begin{subequations}\label{Gamma}
 \begin{align}
\Gamma
&=
-a (a {\hat E} - {\hat L}) 
+
\left \langle a^2 E \cos^2 \theta \right \rangle_\lambda
+
\left \langle \frac{r^2 + a^2}{\Delta} {\hat P} \right \rangle_\lambda\,,
\\ 
\Upsilon^{\phi}
&= - a {\hat E} + {\hat L}
+
\left \langle L \cot^2 \theta \right \rangle_\lambda
+
\left \langle \frac{a}{\Delta} {\hat P} \right \rangle_\lambda\,.
 \end{align}
\end{subequations}
Equations~\eqref{dotQC-ave} and~\eqref{1stlaw-Sago}
can now be substituted into the long-time average of 
the $t$-derivative of Eq.~\eqref{C} given by 
$
\langle {\rmd {\hat C}}/{\rmd t} \rangle_t
=
\langle {\rmd {\hat Q}}/{\rmd t} \rangle_t
- 
2(a {\hat E} - {\hat L}) 
\{
a \langle {\rmd {\hat E}}/{\rmd t} \rangle_t
-
\langle {\rmd {\hat L}}/{\rmd t} \rangle_t
\}
$, 
which easily reveals the equivalence between 
Eqs.~\eqref{dotQ-ave} and~\eqref{dotC-ave}. 

Equations~\eqref{dELdt-ave} and~\eqref{dotQC-ave} are 
the final form of the flux formulae for the energy, azimuthal angular momentum 
and Carter constants (in terms of those for $J_{\alpha}$). 
When the flux formulae of Eq.~\eqref{Flux-nonres} for the non-resonant orbits 
are inserted into Eqs.~\eqref{dELdt-ave} and~\eqref{dotQC-ave}, 
we have the same results obtained by Sago et al.~\cite{Sago:2005fn} 
and displayed in their Eqs.~(3.13),~(3.15),~(3.24), and (3.26), respectively. 
Similarly, the resonant orbits substitution of Eqs.~\eqref{Flux-res} 
in Eqs.~\eqref{dELdt-ave} and~\eqref{dotQC-ave}
gives the same results obtained by Flanagan et al.~\cite{Flanagan:2012kg} 
and displayed in their Eqs. (3.35) --~(3.40), respectively. 
We note, however, that the flux formulae 
of $\langle {\rmd {\hat Q}}/{\rmd t} \rangle_t$ 
in Eq. (3.40) of Ref.~\cite{Flanagan:2012kg} must be supplemented 
by the additional contribution from the symmetric 
interaction Hamiltonian $\Hsym$. 
This is especially clear in view of Eqs.~\eqref{dotQC-ave}:  
in the resonant case, $\langle {\rmd {J}_{r}}/{\rmd t} \rangle_t$ 
and $\langle {\rmd {J}_{\theta}}/{\rmd t} \rangle_t$ 
in Eqs.~\eqref{dotQC-ave} generally require the flux formulae 
of Eq.~\eqref{Flux-res} involving 
${\delta \left\langle \Hsym \right\rangle_\tau}/{\delta w^{\perp}_0}$.

\ack

We thank Leor Barack, Scott Hughes, Maarten van de Meent, Eric Poisson, 
and Adam Pound for useful discussion and feedback on the manuscript. 
SI is particularly grateful to Riccardo Sturani 
for his continuous encouragement. 
SI acknowledges the financial support of JSPS Postdoctoral Fellowship 
for Research Abroad and Ministry of Education - MEC 
during his stay at IIP-Natal-Brazil. 
This work was supported in part by JSPS/MEXT KAKENHI Grant 
No.\ JP16H02183 (R.F.), No.\ JP18H04583 (R.F.), 
No.\ JP17H06358 (H.N., N.S., and T.T.), 
No.\ JP16K05347 (H.N.) and No.\ JP16K05356 (N.S.).

\appendix

\section{Killing vectors and tensors for Kerr geometry}
\label{sec:Killing}

In this Appendix we collect a few key results of Killing vectors 
and tensors of the Kerr spacetime that play a central role 
when constructing constants of motion for Kerr geodesic orbits---
recall Sec.~\ref{subsec:wJ}. 

The stationary and axisymmetric Kerr geometry admits 
the two Killing vectors 
\begin{equation}\label{k-vectors}
t^{\alpha} \equiv \frac{\partial x^{\alpha}}{\partial t}\,,
\quad 
\phi^{\alpha} \equiv \frac{\partial x^{\alpha}}{\partial \phi}\,.
\end{equation}
Also, it is now well known that there is another ``hidden symmetry'' 
of the Kerr geometry associated with the (rank-$2$ irreducible) Killing tensor 
$K_{\mu \nu}  = K_{(\mu \nu)}$~\cite{Carter:1968rr,Carter:1968ks,Walker:1970un} 
that satisfies the Killing equation
$\nabla^{(0)}_{(\lambda} K_{\mu \nu)} = 0$; 
the operator $\nabla^{(0)}_{\mu}$ is the covariant differentiation 
with respect to the Kerr metric $g_{\mu \nu}^{(0)}$, 
and parentheses embracing indices are the total symmetrization 
of a given tensor. 
Explicit expressions for $K_{\mu \nu}$ may need 
a standard Kinnersley null tetrad---recall Eq.~\eqref{Kerr-hojo}: 
\begin{equation}\label{ln-v}
\ell^{\alpha} \equiv 
{\Delta^{-1}}\,\left(r^2 + a^2,\, \Delta,\,0,\,a \right)\,,
\quad
n^{\alpha} \equiv 
\left(2 \Sigma \right)^{-1} 
\left(r^2 + a^2,\,-\Delta,\,0,\,a \right)\,, 
\end{equation}
and 
\begin{equation}\label{m-v}
m^{\alpha} \equiv 
\frac{1}{ \sqrt{2}\sin \theta  (r + i a \cos \theta) } 
\left(i a \sin^2 \theta,\,0,\,\sin \theta,\, i  \right)\,,
\end{equation}
or the associated basis $1$-forms: 
\begin{equation}\label{ln-f}
\ell_{\alpha} = 
{\Delta^{-1}}\,\left(-\Delta,\, \Sigma,\,0,\,a \Delta \sin^2 \theta \right)\,,
\quad
n_{\alpha} = 
\left(2 \Sigma \right)^{-1} 
\left(-\Delta,\,-\Sigma,\,0,\,a \Delta \sin^2 \theta \right)\,, 
\end{equation}
and 
\begin{equation}\label{m-f}
m_{\alpha} = 
\frac{1}{ \sqrt{2}  (r + i a \cos \theta) } 
\left(-i a \sin \theta,\,0,\,\Sigma,\, i (r^2 + a^2) \sin \theta  \right)\, 
\end{equation}
that satisfy $\ell^{\mu} n_{\mu} = -1$ and 
$m^{\mu} {\overline {m}}_{\mu} = 1$ 
(the overbar denotes the complex conjugate).  
Making use of these basis 1-forms, 
we then write the Killing tensor as~\cite{Drasco:2005is,Yasui:2011pr} 
\begin{equation}\label{defK}
K_{\mu \nu} \equiv 
2 a^2\, \cos^2 \theta \, \ell_{(\mu} n_{\nu)} 
+ 2 r^2\,  m_{(\mu} {\overline {m}}_{\nu)}\,.
\end{equation}

The Kerr metric 
of Eq.~\eqref{Kerr} can also be written in terms of the basis 1-forms 
$\{\ell_{\mu},\,n_{\mu},\,m_{\mu},\,{\overline {m}}_{\mu}\}$ as 
\begin{equation}\label{Kerr-K}
g^{(0)}_{\mu \nu} 
= 
-2 \, \ell_{(\mu} n_{\nu)} + 2\, m_{(\mu} {\overline {m}}_{\nu)}\,.
\end{equation}
With Eqs.~\eqref{Kerr-hojo}, this can be substituted into the right-hand side 
of Eq.~\eqref{defK} to establish the ``duality'' of the Killing tensor: 
\begin{equation}\label{K-rth}
K_{\mu \nu} 
= 
2 \Sigma \, \ell_{(\mu} n_{\nu)} + r^2 g^{(0)}_{\mu \nu} 
= 
2 \Sigma \,m_{(\mu} {\overline {m}}_{\nu)} 
- a^2 \cos^2 \theta g^{(0)}_{\mu \nu}\,.
\end{equation}

\section{Teukolsky formalism of black-hole perturbation theory}
\label{sec:Teukolsky}

We include here the essential results of the well-studied
Teukolsky formalism~\cite{Teukolsky:1973ha,Press:1973zz,Teukolsky:1974yv} 
which support our discussion in Sec.~\ref{sec:flux}; 
see also, e.g.,~Refs.~\cite{Sasaki:2003xr,Sago:2005fn,Merlin:2016boc} 
to complement our presentation. 

The master variables in the Teukolsky formalism are 
essentially Weyl scalars $\psi_0$ and $\psi_4$ defined--- 
recall the Kinnersley null tetrad of Eq.~\eqref{ln-v}---by
\begin{align}\label{W-scalar0}
{}_s \Psi \equiv
\begin{cases}
\psi_0 \! &\equiv
-C_{\alpha \beta \gamma \delta} \,
\ell^{\alpha} m^{\beta} \ell^{\gamma} m^{\delta} \,, 
\quad s = 2 \,,
\cr
\rho^{-4} \psi_4 \! &\equiv
-\rho^{-4} C_{\alpha \beta \gamma \delta} \,
n^{\alpha} \conj{m}^{\beta} n^{\gamma} \conj{m}^{\delta} \,,
\quad s = -2 \,,
\end{cases}
\end{align}
where $C_{\alpha \beta \gamma \delta}$ is the (perturbed) Weyl tensor,
and $\rho \equiv (r- i a \cos \theta)^{-1}$. 
The master variables ${}_s \Psi$ satisfy the Teukolsky equation 
\begin{equation}\label{def-Teu0}
{}_s {\cal O} \, {}_s \Psi 
= 4 \pi \Sigma \, {}_s 
\tau_{\alpha \beta}\, T^{\alpha \beta} \,,
\end{equation}
where ${}_s {\cal O}$ and ${}_s \tau_{\alpha \beta}$ are 
differential operators for the spin-weight $s$, 
and $T^{\alpha \beta}$ is the energy-momentum tensor of the matter source 
(point-particle source, etc). 
In the bulk of our paper, we may need the explicit expression 
for the adjoint of ${}_{-2} \tau_{\alpha \beta}$
which is given (see, e.g., Eq.~(A.28) of Ref.~\cite{Sago:2005fn}) by~\footnote{
Suppose a linear differential operator $M$ acts 
on an $n$-index tensor $T$, taking it to another $k$-index tensor 
$M \circ T$. 
The adjoint of $M$ is defined by $M^{\dagger}$ 
so that the relation  
$(M_1 \circ M_2)^{\dagger} = M_2^{\dagger} \circ M_1^{\dagger}$
is satisfied for any such pair of two linear operators
$M_1$ and $M_2$~\cite{Wald:1978vm}. 
Here, we use a dagger $(\dagger)$ to denote the adjoint, 
and a plus $(+)$ to imply the transformation 
$(\omega,\,m) \to (-\omega,\,-m)$. 
Notice that these symbols would differ from the common choices: 
the $\dagger$ and $+$ operators correspond 
to a star $(*)$ and dagger $(\dagger)$ in Ref.~\cite{Sago:2005fn}, 
respectively.}
\begin{align}
\label{def-taud}
 {}_{-2}\tau_{\alpha \beta}^{\dagger} &\equiv 
- \Biggl[ 
\frac{\Sigma}{\sqrt{2}} n_{(\alpha} {m}_{\beta)}
\left(
{\cal D}^{+}_{1} \frac{{\rho}^2}{\conj{\rho}^4} {\cal L}^{+}_{2}
+
{\cal L}^{+}_{2} \frac{{\rho}^2} {\conj {\rho}^4} {\cal D}_{1}^{+}
\right) \Delta \cr 
\quad \quad &+ 
n_\alpha n_\beta
\frac{1}{\conj \rho}
{\cal L}^{+}_{1} \frac{1}{{\conj {\rho}}^4}{\cal L}^{+}_2 
+
\frac{1}{2}m_\alpha m_\beta
\frac{\rho^2}{\conj \rho}
{\cal D}^{+}_{0} \frac{1}{{\conj {\rho}}^4}{\cal D}^{+}_0  \Delta^2 
\Biggr] \conj{\rho}^3 \,, 
\end{align}
where the plus $(+)$ operators are 
\begin{align}\label{def-LDp}
{\cal D}^{+}_n \equiv
- \frac{2 \Sigma}{\Delta} n^{\alpha} \partial_{\alpha} 
+ \frac{2 n (r - M)}{\Delta}\,, 
\quad 
{\cal L}^{+}_s \equiv
 \frac{\sqrt{2}}{\conj{\rho}}{m}^{\alpha} \partial_{\alpha} + s \cot \theta\,,
\end{align}
with integers $n$ and $s$. 
The explicit expressions for other operators are displayed 
in, e.g., Appendix A of Ref.~\cite{Sago:2005fn}.

The Teukolsky equation of Eq.~\eqref{def-Teu0} admits 
a full separation of variables in the frequency domain. 
We may write the solution ${}_s \Psi$
and the source term $ {}_s T \equiv {}_s 
\tau_{\alpha \beta} \, T^{\alpha \beta} $ as 
\begin{subequations}\label{F-Teukolsky}
\begin{align}\label{F-PsiT}
_s\Psi &=
\int_{-\infty}^{\infty} \rmd \omega 
\sum_{\ell = 2}^{+\infty} \sum_{m=-\ell}^{+\ell}
e^{-i \omega t } \,_s R_{\wlm} (r) \,_{s} S_{\wlm} (\theta,\,\phi)\,, \\ 
4\pi\Sigma\,_s {T} &= 
\int_{-\infty}^{\infty} \rmd \omega 
\sum_{\ell = 2}^{+\infty} \sum_{m=-\ell}^{+\ell}
e^{-i \omega t  }
\,_s {T}_{\wlm} (r) \,_{s} S_{\wlm}(\theta,\,\phi)\,,
\end{align}
\end{subequations}
where $\omega$ is a continuous frequency, $\ell$ and $m$ are 
integers, and we have introduced the spin-weighted spheroidal harmonics 
$\,_{s} S_{\wlm} \equiv 
(1/ \sqrt{2 \pi}) \,_{s} \Theta_{\wlm}(\theta) e^{i m \phi}$. 
Substituting Eqs.~\eqref{F-Teukolsky} into Eq.~\eqref{def-Teu0}, 
we obtain the angular and radial Teukolsky equations
\begin{subequations}\label{F-TeukolskyEqs}
\begin{align}\label{Spheroidal}
& \left\{
\frac{1}{\sin \theta} \frac{d}{d \theta} 
\left( \sin \theta \frac{d}{d \theta} \right)
- \,_s U_{\wlm}(\theta) 
\right\}  \,_{s} \Theta_{\wlm}(\theta) = 0 \,, \\
\label{Radial}
& \left\{
\Delta^{-s} \frac{\rmd}{\rmd r} 
\left( \Delta^{s + 1} \frac{\rmd}{\rmd r} \right) 
- \,_s V_{\wlm}(r) 
\right\} \,_s R_{\wlm} (r) = \, _s {T}_{\wlm}\,, 
\end{align}
\end{subequations}
with potentials 
\begin{subequations}\label{def-UV}
\begin{align}
\,_s U_{\wlm}(\theta) &\equiv 
-\, _s \lambda_{\wlm} + \frac{(m + s \cos \theta)^2}{\sin^2 \theta}
+ a^2 \omega^2 \sin^2 \theta 
+ 2 s a \omega \cos \theta - 2 a m \omega -s \,, \\
\,_s V_{\wlm}(r) &\equiv 
\,_s \lambda_{\wlm} 
- 4 i s \omega r
- \frac{K^2 - 2 i s (r - M) K}{\Delta} \,, 
\end{align}
\end{subequations}
and $K \equiv \omega (r^2 + a^2) - a m$.

The differential equation of Eq.~\eqref{Spheroidal} defines 
the (polar part of) the spin-weighted spheroidal harmonics 
$\,_{s} \Theta_{\wlm}(\theta)$ normalized as 
$\int_0^{\pi} \rmd \theta \sin \theta |\,_{s} \Theta_{\wlm}|^2 = 1$, 
and the associated eigenvalues $\,_s \lambda_{\wlm}$ in Eqs.~\eqref{def-UV}. 
At the same time, the homogeneous solution of Eq.~\eqref{Radial} 
defines the four independent ``mode'' functions, 
depending on their boundary conditions at infinity and the horizon. 
In keeping with common nomenclature, they are defined 
(see, e.g., Ref.~\cite{Galtsov:1982hwm}) by 
\begin{subequations}\label{mode1}
\begin{align}\label{eq:Rindef}
\,_s R_{\wlm}^{\rm in}(r_{\ast}) &= 
\left\{ 
\begin{array}{ll} 
\,_s B^{\rm trans}_{\wlm}
\Delta^{-s} e^{- i p_{m \omega} r_{\ast}} 
 & \mbox{ ($r_{\ast} \to -\infty$)\,, }  \\
\,_s B^{\rm inc}_{\wlm} r^{-1} e^{-i \omega r_{\ast}} 
+ 
\,_s B^{\rm ref}_{\wlm} r^{-2s-1} e^{i\omega r_{\ast}}   
 \qquad       &
\mbox{ ($r_{\ast} \to \infty$)\,,}
\end{array} \right. \\
\label{eq:Rupdef} 
\,_s R_{\wlm}^{\rm up}(r_{\ast}) &= 
\left\{ 
\begin{array}{ll}  
  \,_s C^{\rm inc}_{\wlm}   e^{ i p_{m \omega} r_{\ast}} 
  + \,_s C^{\rm ref}_{\wlm} \Delta^{-s} e^{-i p_{m \omega} r_{\ast}}
  & \mbox{ ($r_{\ast} \to -\infty$)\,, } \\
\,_s C^{\rm trans}_{\wlm} r^{-2s-1} e^{i \omega r_{\ast}}
& \mbox{ ($r_{\ast} \to \infty$)\,, }
\end{array} \right.
\end{align}
\end{subequations}
for the ``up'' and ``in'' modes, and 
\begin{subequations}\label{mode2}
\begin{align}
\,_s R^{{\rm out}}_{\wlm} (r_{\ast})  
&= \Delta^{-s} {\overline {\,_{-s} R^{{\rm in} }_{\wlm}} (r_{\ast})}\,, \\
\,_s R^{{\rm down}}_{\wlm} (r_{\ast}) 
&= \Delta^{-s} {\overline {\,_{-s} R^{{\rm up} }_{\wlm}} (r_{\ast})}\,,
\end{align}
\end{subequations}
for the ``out'' and ``down'' modes, where 
$p_{m \omega} \equiv \omega - m a / (2 M r_{+} )$ 
and $r_{\ast}$ is the tortoise coordinate that satisfies 
$d r_{\ast}/ d r = (r^2 + a^2)/\Delta$. 
The complex-valued coefficients 
$\,_s B^{\rm inc}_{\wlm},\,\,_s B^{\rm ref}_{\wlm}$ 
and $\,_s B^{\rm trans}_{\wlm}$ 
($\,_s C^{\rm inc}_{\wlm},\,\,_s C^{\rm ref}_{\wlm}$ 
and $\,_s C^{\rm trans}_{\wlm}$ )
are, respectively, incidence, reflection and transmission coefficients 
of the ``in''-mode (``up''-mode) solutions.

For given boundary conditions, the Green's function 
of the Teukolsky equation of Eq.~\eqref{def-Teu0} is defined by 
the solution of the differential equation 
\begin{equation}\label{Green-T}
\,_s {\cal O} \,{\cal G}(x,x') 
= \Delta^{-s} {\delta^{(4)} ( x - x' )}\,,
\end{equation} 
where $\delta^{(4)} (x - x')$ is a $4$D coordinate delta function, 
and its explicit expressions can be constructed 
in terms of the mode functions of Eqs.~\eqref{mode1} and~\eqref{mode2}. 
For example, the retarded Green's function that satisfies 
the retarded boundary condition ${\cal G}_{+}(x,x') = 0$ 
for $t < t'$ is given  (see, e.g., Eq.(A.38) of Ref.~\cite{Sago:2005fn}) by 
\begin{align}\label{Gret-Teukolsky}
\,_{s}{\cal G}_{+}(x,x') &= 
\frac{1}{4 \pi i} \int^{\infty}_{-\infty} \rmd \omega 
\sum_{\ell = 2}^{\infty} \sum_{m=-\ell}^{\ell} 
\frac{e^{- i \omega (t - t')}}
{\omega\, {}_s B^{\rm inc}_{\wlm} \,_s C^{\rm trans}_{\wlm}}
\,_{s} S_{\wlm}(\theta, \phi)
\overline {_{s} S_{\wlm}}(\theta', \phi')
\cr
&\quad 
\times\bigg\{
\,_{s} R^{\rm up}_{\wlm}(r) \,_{s}R^{{\rm in}}_{\wlm}(r') H(r-r')
+
\,_{s} R^{{\rm in}}_{\wlm}(r) \,_{s}R^{\rm up}_{\wlm}(r') H(r'-r)
\bigg\} \,, 
\end{align}
where we have introduced a step function 
$H(x) \equiv \int_{-\infty}^{x} \delta (y) \rmd y$. 
The advanced Green's function ${\cal G}_{-}(x,x')$ 
that satisfies the advanced boundary condition ${\cal G}_{-}(x,x') = 0$ 
for $t > t'$ is similarly obtained by Eq.~\eqref{Gret-Teukolsky} 
with the substitutions, 
$\,_{s} R^{\mathrm {in/up}}_{\wlm} \to \,_{s} R^{\mathrm {out/down}}_{\wlm}$ 
and 
$ \,_s B^{\rm inc}_{\wlm} \,_s C^{\rm trans}_{\wlm} \to 
-  
{\overline {\,_{-s} B^{\rm inc}_{\wlm}}}\, 
{\overline {\,_{-s} C^{\rm trans}_{\wlm}}}$ 
---recall Eqs.~\eqref{mode2}.
Then we can show that the radiative Green's function 
${\cal G}^{\mathrm {(rad)}}(x,x')$ is expressed 
(see, e.g., Eq.(A.43) of Ref.~\cite{Sago:2005fn}) as 
\begin{align}\label{Grad-Teukolsky}
\,_{s}{\cal G}^{\mathrm {(rad)}}(x,\,x') 
&\equiv 
\frac{1}{2}
\left(
\,_{s}{\cal G}_{+}(x,\,x') 
- 
\,_{s}{\cal G}_{-}(x,\,x')
\right) \cr 
&=
\frac{1}{2 i \omega^3} \int^{\infty}_{-\infty} \rmd \omega 
\sum_{\ell = 2}^{\infty} \sum_{m=-\ell}^{\ell}
{e^{- i \omega (t - t')}}
\,_{s} S_{\wlm}(\theta,\, \phi)
\overline {_{s} S_{\wlm}}(\theta',\, \phi')
\cr
&\quad 
\times \bigg\{
\,_{s} {\cal A} \,
\,_{s} R^{\rm down}_{\wlm}(r) {\overline {\,_{-s} R^{{\rm down}}_{\wlm}}}(r') 
+ 
\frac{\omega}{p_{\omega m}} \,_{s} {\cal B} \, 
\,_{s} R^{\rm out}_{\wlm}(r) {\overline {\,_{-s} R^{{\rm out}}_{\wlm}}}(r') 
\bigg\} \,.
\end{align}
We do not need the explicit expressions 
for the normalization factors 
${\,_{s} {\cal A}}$ and ${\,_{s} {\cal B}}$ here, 
but they can be straightforwardly computed 
from the results in Appendix A of Ref.~\cite{Sago:2005fn}. 

Notice that, unlike Eq.~\eqref{Gret-Teukolsky}, 
the radiative Green's function of Eq.~\eqref{Grad-Teukolsky} 
contains \textit{no step function}. 
The step functions in $\,_{s}{\cal G}_{+}(x,x')$ and 
$\,_{s}{\cal G}_{-}(x,x')$ exactly cancel each other out, 
and $\,_{s}{\cal G}^{\mathrm {(rad)}}(x,\,x')$ indeed satisfies 
the homogeneous Teukolsky equation, 
$\,_s {\cal O} \,\,_{s}{\cal G}^{\mathrm {(rad)}}= 0$. 
This is the key property of $\,_{s}{\cal G}^{\mathrm {(rad)}}(x,x')$ 
that allows us to construct the radiative interaction Hamiltonian 
in the simple ``homogeneous'' form of Eq.~\eqref{F-Hrad}, 
leading to the flux formulae displayed in 
Eqs.~\eqref{Flux-nonres} and~\eqref{Flux-res}; 
for the full details for obtaining Eq.~\eqref{F-Hrad} 
from Eq.~\eqref{Grad-Teukolsky}, once again we refer readers 
to Appendix A of Ref.~\cite{Sago:2005fn}.

\section{More on evolution of constants of motion}
\label{sec:dQdt-GSF}

Unlike the flux formulae of Eqs.~\eqref{Flux-nonres} and~\eqref{Flux-res}, 
it is not so difficult to express the evolution of the constants of motion 
${\dot P}_{\alpha}$---see Eq.~\eqref{def-P}---directly in terms of 
the local self-force $f_{\mu} (\sim O(\eta))$~\cite{Mino:1996nk,Quinn:1996am}. 
It is given in our notation by using Refs.~\cite{Ori:1997be,Sago:2005fn}:~\footnote{
To be precise, the proper time of those literature 
is ${\tilde \tau}$ normalized with respect to the background Kerr metric 
$g_{\mu \nu}^{(0)}$, which differs from ours $\tau$ of Eq.~\eqref{norm-xu}. 
However, this difference is negligible here because 
$\rmd \tau = \rmd {\tilde \tau} (1 + \Hint)$~\cite{Sago:2008id}, 
and the self-force is already $f_{\alpha} \sim O(\eta)$.
}
\begin{equation}\label{dPdtau-f}
{\dot P}_{\alpha} 
= \left(\frac{\partial P_{\alpha}}{\partial u_{\mu}} \right)_{\!x} 
\,f_{\mu} + O(\eta^2)\,,
\end{equation}
this is the so-called ``forcing term'' in Refs.~\cite{Hinderer:2008dm,Gair:2010iv,vandeMeent:2013sza,vandeMeent:2018rms}. 
The objective of this appendix is to examine 
how this expression can be derived in the Hamiltonian formulation. 

We write Hamilton's equations in the ``mixed'' canonical variables of 
$(x^\mu,\,P_\alpha) \equiv (x^\mu(X,\,P),\,P_\alpha)$. 
From the Hamilton equation for $P_\alpha$, we quickly obtain 
\begin{align}\label{dPdtau-Sago}
 {\dot {P}_{\alpha}}
 &= 
 -\left( \frac{\partial H }{\partial {X}^{\alpha}} \right)_P \cr
 &= - 
 \left(
 \frac{\partial x^{\mu}}{\partial {X}^{\alpha}} 
 \frac{\partial H(x,\,P;\,\gamma)}{\partial x^{\mu}} \right)_{P} \cr
 &= -
 \left(
 \frac{\partial P_{\alpha}}{\partial u_{\mu}}
 \right)_{x}
 \left(
 \frac{\partial \Hint (x,\,P;\,\gamma) }{\partial x^{\mu}}
 \right)_{\!P} \,,
\end{align}
where $\partial P_{\alpha} / \partial X^{\beta} = 0$,  
and the last equality follows from 
$(\partial H^{(0)}(P) / \partial X^{\alpha})_P = 0$ 
as well as the identity 
\begin{equation}\label{Liouville}
\left(
\frac{\partial P_{\alpha}}{\partial u_{\mu} }
\right)_{\!x} 
= \{x^\mu ,\,P_\alpha \}_{x,u}
= \{x^\mu ,\,P_\alpha \}_{X,P}
=
\left(
\frac{\partial x^{\mu}}{\partial X^{\alpha}}
\right)_{\!P}\,,
\end{equation}
with the standard Poisson bracket $\{x^\mu,\,P_\alpha \}$~\footnote{
Given two functions $f(x,\,u)$ and $g(x,\,u)$ 
in the canonical variables $(x^{\mu},\,u_{\mu})$ 
on the phase space, the Poisson bracket takes the form 
$\{f,\,g \}_{x,u} \equiv 
(\partial f / \partial x^{\mu})(\partial g / \partial u_{\mu})
-
(\partial g / \partial x^{\mu})(\partial f / \partial u_{\mu})$.
}.
This gives a derivation of $\dot P_{\alpha}$ expressed 
in the form of Eq.~\eqref{dPdtau-f} in the language of Hamiltonian formalism. 
In particular, the comparison of Eq.~\eqref{dPdtau-f} 
with Eq.~\eqref{dPdtau-Sago} reveals an interesting relation,
\begin{equation}\label{F_a}
f_{\mu} 
= 
\left. -\left(
\frac{\partial \Hint}{\partial x^{\mu}}
\right)_{P} \right|_\gamma \,.
\end{equation}

The flux formulae of Eqs.~\eqref{dELdt-ave} and~\eqref{dotQC-ave}--- after 
inserting Eqs.~\eqref{Flux-nonres} or~\eqref{Flux-res} into them)--- 
are mathematically identical to the radiative part 
of the Hamilton equations of Eqs.~\eqref{dPdtau-f} 
and~\eqref{dPdtau-Sago} after the long-time average with respect to $\tau$.
Thus, their comparison provides an important test 
for the numerical self-force calculation as well as the foundation 
of the flux formulae (i.e., Hamiltonian formulation). 
Such a comparison has been extensively carried out 
for a variety of orbital configurations 
when the relevant self-force codes become available~\cite{Barack:2007tm,Barack:2010tm,vandeMeent:2016pee,vandeMeent:2017bcc}, 
and we have found perfect agreement (within numerical accuracy).



\end{document}